%% file: main1127.tex
\documentclass[10pt,journal]{IEEEtran}

\usepackage{siunitx} 
\usepackage[final]{graphicx} 
\usepackage{amsmath} 
\usepackage{amssymb,times}
\usepackage{subfigure}
\usepackage{cite}
\usepackage{algorithm}
\usepackage{algorithmic}
\usepackage{epstopdf}
\usepackage{color}
\usepackage{url}
\newcommand\undermat[2]{%
  \makebox[0pt][l]{$\smash{\underbrace{\phantom{%
    \begin{matrix}#2\end{matrix}}}_{\text{$#1$}}}$}#2}
\usepackage{hyperref}
\input{common}

\begin{document}
\title{Joint Waveform and Receiver Design for Co-Channel Hybrid Active-Passive Sensing with Timing Uncertainty
\thanks{This work was supported in part by the National Science Foundation
under grants ECCS-1609393 and ECCS-1923739.}
\thanks{Corresponding author: Hongbin Li.}
}
\author{Fangzhou Wang,~\IEEEmembership{Student Member,~IEEE,} and
 Hongbin Li,~\IEEEmembership{Fellow,~IEEE,}
\thanks{F. Wang and H. Li are with the Department of Electrical and Computer Engineering, Stevens Institute of Technology, Hoboken, NJ 07030 USA (e-mail: fwang11@stevens.edu; hli@stevens.edu)}
}

\maketitle

\begin{abstract}
We consider a hybrid active-passive radar system that employs a wireless source as a passive illuminator of opportunity (IO) and a co-channel active radar transmitter operating in the same frequency band to seek spectral efficiency. The hybrid system can take advantage of the strengths of passive radar (e.g., energy efficiency, bi-/multi-static configuration, and spatial diversity) as well as those of active radar (dedicated transmitter, flexible transmit beam steering, waveform optimized for sensing, etc.). To mitigate the mutual interference and location-induced timing uncertainty between the radar and communication signals, we propose two designs for the joint optimization of the radar waveform and receive filters. The first is a max-min (MM) criterion that optimizes a worst-case performance metric over a timing uncertainty interval, and the other a weighted-sum (WS) criterion that forms a weighted sum of the performance metric at each delay within the delay uncertainty interval. Both design criteria result in nonconvex constrained optimization problems that are solved by sequential convex programming methods. When timing uncertainty vanishes, the two designs become identical and admit a simpler solution. Numerical results are presented to demonstrate the performance of the proposed hybrid schemes in comparison with conventional active-only and passive-only radar systems.
\end{abstract}

\begin{IEEEkeywords}
Hybrid active-passive radar, joint waveform and receiver design, timing uncertainty, nonconvex optimization
\end{IEEEkeywords}

\IEEEpeerreviewmaketitle

\section{Introduction}
\label{sec:introduction}
Passive radar is not equipped with a dedicated transmitter but uses existing wireless sources as illuminators of opportunity (IOs) to detect and track targets of interest \cite{GriffithsBaker05,PalmerDavis13,GogineniRangaswamy14,ZhangAmin16,HackHimedSaville14,ZhangLiHimed16,WangLi18,KarthikBlum18,LiHeBlum19,ChenBlum19}. It has a number of advantages when compared with its active counterpart. First, many IO sources are usually available, including frequency modulation (FM) radio, television, digital audio/video broadcasting, cellular systems, satellites, and many others \cite{GriffithsBaker05,PalmerDavis13,GogineniRangaswamy14,ZhangAmin16}. Second, passive radar is more covert and economical since it does not require a transmitter and there is no transmitter operation related cost. Finally, the bi-static or multi-static configuration of passive radar enables it to exploit the associated spatial diversity. Despite such advantages, passive radar cannot apply beam steering on transmit routinely used in active radar. It has no control on the transmitter location/power and, moreover, the IO waveform is intended for communication, not optimized for sensing purpose.

To address the above limitations, a new paradigm of hybrid radar that consists of both an active and a passive component was introduced in \cite{GaoHimed17}. The motivation was to take advantage of the strengths of both active and passive sensing. On one hand, the passive component allows the hybrid system to easily form a bi-static or multi-static configuration with existing IOs to obtain spatial diversity, while reducing the transmit power of the active component by leveraging the illumination from the IOs. On the other hand, the active component of the system can steer its transmit beam as needed and employ optimized transmit waveform to complement the passive component. The hybrid system \cite{GaoHimed17} assumes the passive and active components operate in different frequency bands with no mutual interference. In recent years, there is an increasing interest in co-channel existence of radar and communication systems due to spectrum shortage (e.g., \cite{LiPetropuluTrappe16,ZhengLopsGrossi18,WangLiGovoni19,ZhengLopsWang2019,MishraKoivunen2019}). There have been a number of efforts exploring co-channel illuminations from both an active radar and communication sources to perform radar functions \cite{BicaKoivunen15,BicaKoivunen17,HeBlum2019}, effectively resulting in co-channel hybrid active-passive radars. Specifically, \cite{BicaKoivunen15} examined radar waveform optimization based on the Neyman-Pearson detection performance and mutual information by exploiting the communication signals scattered off the target as useful target information at the radar receiver. Meanwhile, \cite{BicaKoivunen17} proposed a delay estimator from a multicarrier radar that exploits the communication signal in a passive manner. Recently, the Cramer-Rao bound (CRB) for target localization was derived in \cite{HeBlum2019} to show the benefit of utilizing the target returns contributed from the communication transmitters in a co-channel multi-input multi-output (MIMO) radar and MIMO communication system.

A major challenge of a co-channel hybrid active-passive radar is mutual interference. Specifically, the radar receiver observes a target echo due to the active source and, simultaneously, another one due to the passive IO. While each alone is useful, observed together, they also interfere with one another, which has to be addressed by careful transmitter/waveform and receiver design. The subject has been under intensive investigation for active radar, including transmit waveform designs to meet certain requirements on the temporal or spatial characteristics of the radar waveform (e.g., \cite{MaioNicola09,MaioFarina11,CuiFogliaLi17,YuCuiKongLi19}), as well as joint transmit waveform and receiver designs with the additional goal to mitigate clutter or mutual interference created by different transmitters in MIMO radars \cite{AubryDeMaioFarina13,NaghshStoicaAubry14,AubryDeMaioNaghsh15,CiuonzoMaio2015,ChengAubryMaio17,ZhaoPalomar17,CuiCarotenutoKong17,CuiFuYuLi18}. Such designs often utilize prior knowledge of the target and/or clutter environment, and fall within the general framework of cognitive or fully adaptive radar \cite{Haykin06,BellRangaswamy141,BellRangaswamy142}. However, the design problem for hybrid radar is different, as the design parameters at disposal are those related to the active transmitter and receiver, e.g., the radar waveform and receive filters implemented at the receiver of the hybrid radar, and in general we cannot change the waveform and parameters of IO.

In this paper, we examine the joint design of radar waveform and receive filters for a hybrid active-passive radar system which consists of one IO and a monostatic active radar operating in the same frequency band. We consider a scenario where the interference from the radar to the communication system is negligible due to directive transmission of the radar \cite{WangJohnsonBaker17,NTIA10,sanders2012analysis}. We assume the IO waveform is partially known, i.e., the modulation format and pulse shaping waveform is known but the information symbols are unknown due to privacy related issues. Such knowledge was employed in passive radar to improve the target detection and estimation performance \cite{KarthikBlum18,ChenBlum19}. Here, we use it for the joint design of the radar waveform and receive filters to help separating the target echoes from the active and passive illumination and minimizing mutual interference. Since there may be uncertainties related to the locations of the target and/or the IO, which can cause an uncertainty in the timing of the radar and communication signals, we introduce two design criteria to cope with this issue. The first is a max-min (MM) criterion, which tries to optimize the worst-case performance metric, i.e., the output signal-to-interference-plus-noise ratio (SINR), within a delay uncertainty interval. The second is a weighted-sum (WS) criterion that employs a set of weights to form a weighted sum of the SINR at each delay within the delay uncertainty interval, which is useful in applications with prior knowledge about the timing uncertainty. Both design criteria result in nonconvex constrained optimization problems. We propose sequential convex programming methods to iteratively optimize the radar waveform and receive filters. In the absence of timing uncertainty, we show the two designs become identical, leading to a simplified solution. Our extensive numerical results show the proposed hybrid designs can achieve considerable performance improvements over conventional active-only or passive-only system.

The remainder of the paper is organized as follows. The problem of interest is formulated in \ref{sec:signalmodel}. The timing uncertainty and proposed design criteria are discussed in Section \ref{subsec:ASynchronized}. Solutions to the proposed designs are presented in Section \ref{sec:proposedmethod}. Section \ref{sec:simulationresults} contains numerical results and discussion, followed by conclusions in Section \ref{sec:conclusion}.

\emph{Notations}: We use boldface symbols for
    vectors (lower case) and matrices (upper case). $\Cset^{N\times1}$
    denotes the set of $N\times1$ vectors of complex numbers,
    $(\cdot)^T$ the transpose, and  $(\cdot)^H$ the conjugate
    transpose. $\mathbf{0}$ and $\Ibf$ denote a matrix with zero
entries and an identity matrix, respectively. $\mathbb{E}\{\cdot\}$
represents the statistical expectation. $\mathbb{N}_K$ denotes a set defined as $\{-K,\cdots,0,\cdots,K\}$. $\text{tr}(\cdot)$ is the matrix trace operator.  $\mathcal{N}(\ubf,\Sigmabf)$ denotes the Gaussian distribution with mean $\ubf$ and covariance matrix $\Sigmabf$. $\Abf\succeq\mathbf{0}$ indicates that $\Abf$ is a positive semi-definite matrix. Finally, $\mathcal{O}(\cdot)$ denotes the Landau notation for complexity.
\section{Signal Model and Problem Formulation}
\label{sec:signalmodel}

\begin{figure}[t]
\centering
\includegraphics[width=2.5in]{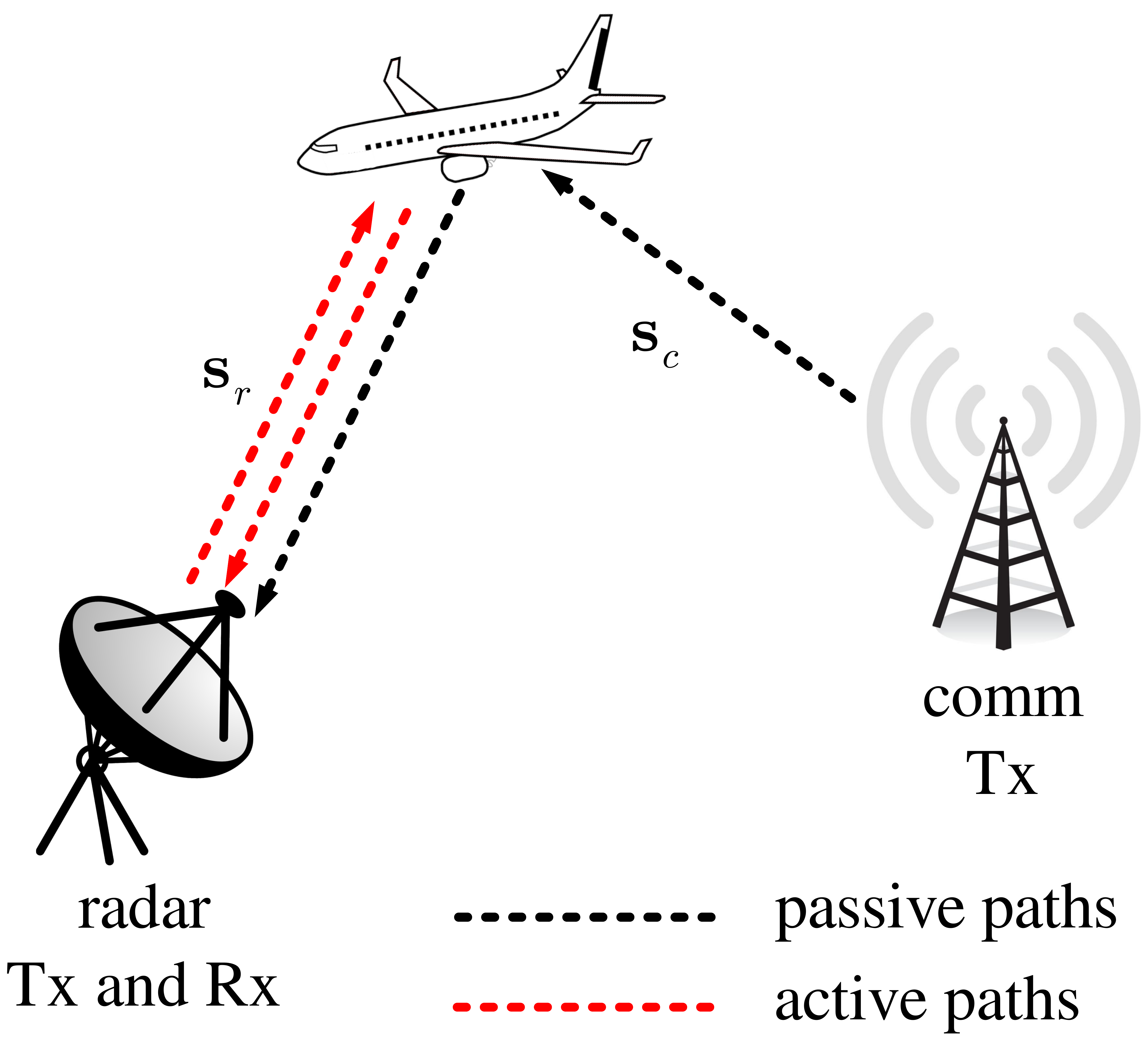}
\caption{A hybrid radar system with an active transmitter, an IO, and a shared receiver that receives from both the active and passive paths.}
\label{fgr:systemconfiguration}
\end{figure}
In this paper, we consider a hybrid active-passive radar system as shown in Fig.\,\ref{fgr:systemconfiguration}, where a broadcasting communication transmitter (Tx), referred to as an illuminator of opportunity (IO), and a monostatic radar system in the same neighborhood operate in the same frequency band. The radar system also tries to leverage the IO as a passive Tx, which is utilized along with the radar's own Tx to illuminate a target. In the following, we first introduce the signal model for the hybrid system and then discuss the formulation of the design.

For the hybrid system, the radar's active Tx emits a probing waveform, and its receiver (Rx) receives target echoes of both the probing waveform and the IO waveform simultaneously. We can obtain a baseband equivalent discrete-time signal by first down-converting the received signal and passing it through a sampling scheme with a sampling interval $T_s$. We assume that $T_s$ is very small so we obtain a good approximation of the continuous-time signal in an observation window $NT_s$. Then, the noise corrupted received signal $\ybf$ can be written as
\ben\label{equ:sync}
\ybf=\alpha_r\sbf_r+\alpha_c\sbf_c+\wbf,
\een
where $\sbf_r\in\Cset^{N\times1}$ and $\sbf_c\in\Cset^{N\times1}$ denote the sampled waveform vector of the active radar, and respectively, the IO signal, $\alpha_r$ and $\alpha_c$ are the amplitudes of the target due to the active probing signal and the passive IO signal, respectively, and $\wbf$ denotes the noise with zero mean and covariance matrix $\sigma^2\Ibf$. Note that the target amplitudes can be expressed as
\ben
\alpha_r=\rho h_r,\qquad
\alpha_c=\rho h_c,
\label{equ:rcs}
\een
where $\rho$ denotes radar cross section (RCS) of the target, while $h_r$ and $h_c$ denote channel coefficients that include the effects of antenna gain and propagation related attenuation.

The received signal is filtered through two filters $\wbf_r$ and $\wbf_c$, which are used to separate the reflections from the radar Tx and communication Tx:
\ben\label{equ:mfout}
\begin{cases}
  y_r=\wbf_r^H\ybf=\alpha_r\wbf_r^H\sbf_r+\alpha_c\wbf_r^H\sbf_c+\wbf_r^H\wbf,\\
  y_c=\wbf_c^H\ybf=\alpha_r\wbf_c^H\sbf_r+\alpha_c\wbf_c^H\sbf_c+\wbf_c^H\wbf.
\end{cases}
\een

\emph{Remark 1}: In this paper, we focus on a single-pulse scenario, where the target Doppler effect is neglected. For moving target detection, the radar can transmit a pulse train, i.e., multiple copies of $\sbf_r$, over a coherent processing interval (CPI). Each pulse is first filtered by $\wbf_r$ and $\wbf_c$, and then the filtered outputs are processed by Doppler processing to expose the target Doppler shift.

After receive filtering, the output SINR can be expressed as
\ben
\label{equ:SINRDef}
\text{SINR}(\sbf_r,\wbf_r,\wbf_c)=\text{SINR}_{r}(\sbf_r,\wbf_r)+\text{SINR}_{c}(\sbf_r,\wbf_c),
\een
where
\ben
\text{SINR}_{r}(\sbf_r,\wbf_r)=\frac{\mathbb{E}\{\vert\alpha_r\vert^2\}\wbf_r^H\sbf_r\sbf_r^H\wbf_r}{\mathbb{E}\{\vert\alpha_c\vert^2\}\wbf_r^H\mathbb{E}\{\sbf_c\sbf_c^H\}\wbf_r+\sigma^2\wbf_r^H\wbf_r},
\een
and
\ben
\text{SINR}_{c}(\sbf_r,\wbf_c)=\frac{\mathbb{E}\{\vert\alpha_c\vert^2\}\wbf_c^H\mathbb{E}\{\sbf_c\sbf_c^H\}\wbf_c}{\mathbb{E}\{\vert\alpha_r\vert^2\}\wbf_c^H\sbf_r\sbf_r^H\wbf_c+\sigma^2\wbf_c^H\wbf_c},
\een
where the communication waveform $\sbf_c$ is modeled as a random vector because of the message symbols it carries, which are assumed unknown. To facilitate discussions, we define the \emph{channel signal-to-noise ratio} (SNR) as follows:
\ben\label{equ:gammar_r}
\gamma_r=\frac{\mathbb{E}\{\vert\alpha_r\vert^2\}}{\sigma^2}=\rho^2\frac{\mathbb{E}\{\vert h_r\vert^2\}}{\sigma^2},
\een
 and
\ben\label{equ:gammar_c}
\gamma_c=\frac{\mathbb{E}\{\vert\alpha_c\vert^2\}}{\sigma^2}=\rho^2\frac{\mathbb{E}\{\vert h_c\vert^2\}}{\sigma^2}.
\een
Note these quantities are called channel SNRs since they do not include the transmitter power. Given these definitions, \eqref{equ:SINRDef} can be rewritten as
\ben\label{equ:SINRdef}
\begin{split}
\text{SINR}(\sbf_r,\wbf_r,\wbf_c)&=\frac{\gamma_r\wbf_r^H\sbf_r\sbf_r^H\wbf_r}{\gamma_c\wbf_r^H\mathbb{E}\{\sbf_c\sbf_c^H\}\wbf_r+\wbf_r^H\wbf_r}\\
&+\frac{\gamma_c\wbf_c^H\mathbb{E}\{\sbf_c\sbf_c^H\}\wbf_c}{\gamma_r\wbf_c^H\sbf_r\sbf_r^H\wbf_c+\wbf_c^H\wbf_c}.
\end{split}
\een
In this paper, we consider the joint design of the active radar waveform $\sbf_r$ and the receive filters, $\wbf_r$ and $\wbf_c$, by maximizing the output SINR.

\emph{Remark 2}: For the design problem, the communication signal $\sbf_c$ is not fully known because the IO is non-cooperative and the message symbols carried by the communication signal may be unknown to the radar. On the other hand, the modulation format/waveform of the communication signal is usually known, which can be exploited to help the design. However, there are other uncertainties related to the locations of the target and/or the IO, which can cause an uncertainty in the timing of the radar and communication signals. Such uncertainties have to be addressed in the design.

\emph{Remark 3}: In our design, we consider primarily the mutual interference observed at the radar Rx, while the interference created by the radar waveform to a communication Rx is neglected. This reflects scenarios such as spectrum sharing between the air traffic control or marine radar and wireless systems \cite{WangJohnsonBaker17,NTIA10,sanders2012analysis}, in which the radar-to-communication interference is small due to directive transmission of the radar Tx.
\section{Timing Uncertainty and Design Criteria}
\label{subsec:ASynchronized}
In this section, we introduce the design criterion when the location-induced timing uncertainty is absent or present. The timing uncertainty refers to an uncertainty of the relative timing between the communication and radar waveforms, which has an impact on the composition of $\sbf_c$ in \eqref{equ:sync}. To elaborate, we first derive a linear representation for $\sbf_c$ assuming linear modulation for the communication signal, and then discuss how the timing uncertainty affect the representation and design criterion.

With linear digital modulation the baseband communication signal $s_c(t)$ can be expressed as \cite{proakis2001digital}
\ben\label{equ:ldm}
s_c(t)=\sum_{\ell=0}^{L-1}b_{\ell}g(t-\ell T_c),
\een
where $b_{\ell}$ denotes the $\ell$-th communication symbol, $g(t)$ is the symbol pulse shaping function which is assumed to have a duration of $IT_c$, $L$ is the number of communication symbols, and $T_c$ denotes the symbol duration. Here, to be consistent with the discrete model in Section \ref{sec:signalmodel}, we also obtain a discrete-time representation of $s_c(t)$ by using a fine sampling interval $T_s$. Let $P=T_c/T_s$ which is assumed to be an integer. Then, the discrete model of \eqref{equ:ldm} can be rewritten as
\ben\label{equ:ldm_dis}
s_c(pT_s)=\sum_{\ell=0}^{L-1}b_{\ell}g(pT_s-\ell T_c),\,p=0,\cdots,M-1,
\een
where
\ben
M=(L+I-1)P
\een
denotes the total number of samples used to represent the communication waveform. Let $\sbf=[s_c(0),s_c(T_s),\cdots,s_c((M-1)T_s)]^T$. Then, we can express $\sbf$ as \cite{KarthikBlum18}
\ben\label{equ:ss}
\sbf=\Hbf\bbf,
\een
where $\bbf=[b_0,b_1,\cdots,b_{L-1}]^T$ contains the unknown communication symbols and $\Hbf$ is an $M\times L$ waveform matrix:
\begin{equation}\label{equ:Hmatrix}
\Hbf=
\left (
\begin{array}{cccccc}
\gbf_0  & \mathbf{0}_{P\times1} & \cdots & \mathbf{0}_{P\times1}\\
\vdots & \gbf_0  & \ddots & \vdots \\
\gbf_{I-1} & \vdots & \ddots & \mathbf{0}_{P\times1}\\
\mathbf{0}_{P\times1}& \gbf_{I-1}& \ddots & \gbf_0\\
\vdots & \vdots & \ddots & \vdots\\
\mathbf{0}_{P\times1} &\mathbf{0}_{P\times1} & \cdots & \gbf_{I-1}
\end{array}
\right ),
\end{equation}
with $\gbf_i=[g(-iT_c),g(T_s-iT_c),\cdots,g((P-1)T_s-iT_c)]^T$, $i=0,\cdots,I-1$.

\begin{figure}[t]
\centering
\includegraphics[width=3.4in]{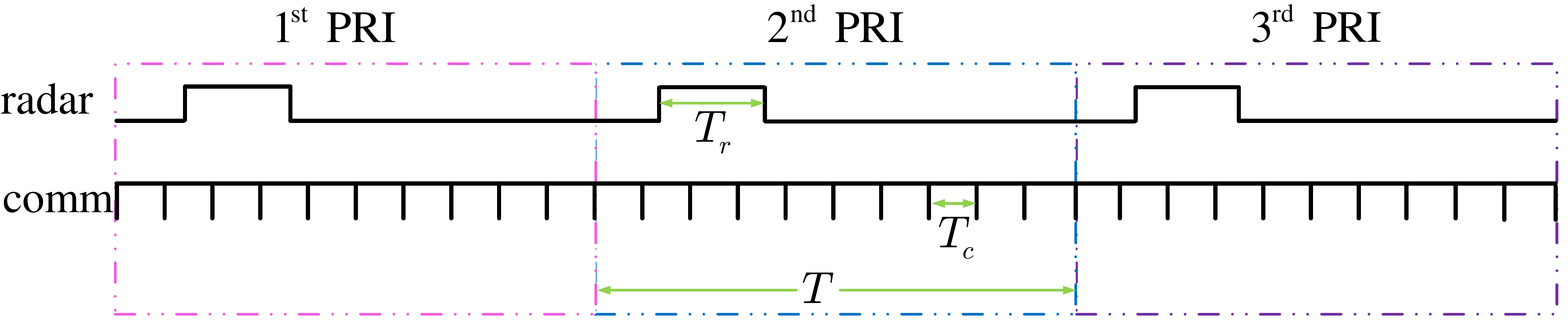}
\caption{An illustration of the relative timing of the received radar and communication waveforms. Note that the radar pulse duration $T_r$ can be smaller or larger than the communication symbol duration $T_c$.}
\label{fgr:systemconfiguration_1}
\end{figure}

We assume the radar system employs a pulse structure with a pulse repetition interval (PRI) $T$ and there are $L$ communication symbols in one radar PRI, i.e., $T=LT_c$. Each radar pulse has a duration $T_r$, which is also the observation interval in \eqref{equ:sync}, i.e., $T_r=NT_s$. The relative relation of $T$, $T_r$, and $T_c$ is illustrated in Fig.\,\ref{fgr:systemconfiguration_1}, which shows that the communication signal $\sbf_c$ observed in \eqref{equ:sync} is one segment of $\sbf$. However, the exact position, i.e., the starting and ending samples of $\sbf_c$, within $\sbf$ depends on the knowledge of the arrival time of the target returns, which are discussed next.
\subsection{No Timing Uncertainty}
\label{subsubsec:WitouthUncertainty}
\begin{figure}[t]
\centering
\includegraphics[width=2.5in]{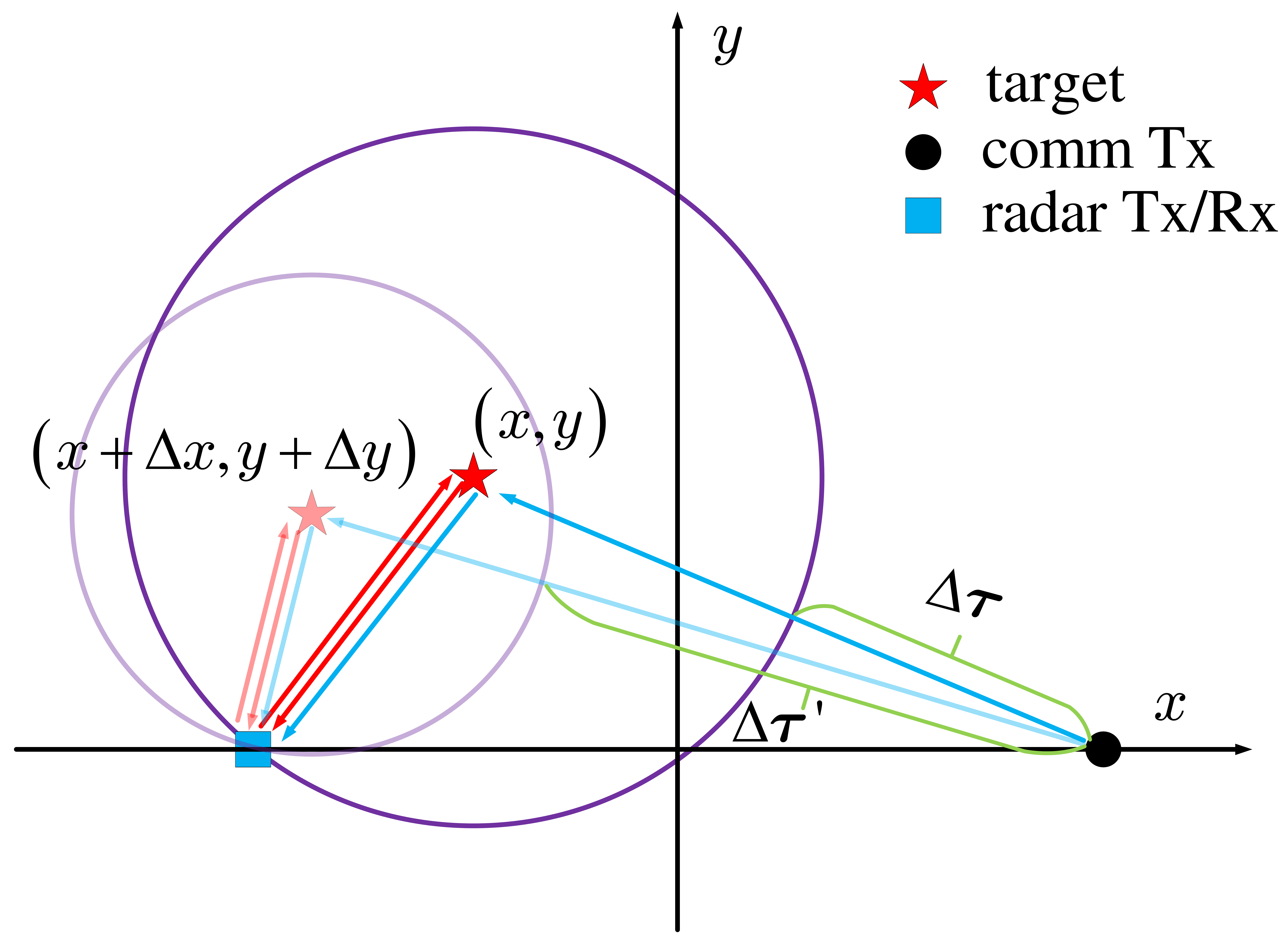}
\caption{Illustration of timing uncertainty caused by location uncertainty.}
\label{fgr:systemconfiguration_2}
\end{figure}

Let $\tau_r$ denote the propagation delay of the active path, i.e., the delay due to the transmission from the radar Tx to the target and then from the target back to the radar Rx (active path), and $\tau_c$ be similarly defined for the passive path. Suppose there is no timing uncertainty such that $\tau_r$ and $\tau_c$ are accurately known. Let $(x,~y)$, $(x_r,~y_r)$, and $(x_c,~y_c)$ denote the coordinates of the target, radar, and IO, respectively, in a two-dimensional Cartesian coordinate system. Then, the propagation delay difference $\Delta\tau=\tau_r-\tau_c$ as shown in Fig.\,\ref{fgr:systemconfiguration_2} can be represented as
\ben
\Delta\tau=\frac{\sqrt{(x-x_r)^2+(y-y_r)^2}-\sqrt{(x-x_c)^2+(y-y_c)^2}}{c},
\een
where $c$ is the speed of light. Positive $\Delta \tau$ indicates that the IO is closer to the target than the active Tx and vice versa.

Note that $\Delta\tau$ is used by the receiver to determine the relative position of $\sbf_c$ within $\sbf$. Without loss of generality, we assume when timing uncertainty is absent, $\sbf_c$ is located in the middle of $\sbf$, which is referred to as the nominal position. When timing uncertainty is present (considered in the next subsection), the real position of $\sbf_c$ will be shifted to either the left or the right of the nominal position. It follows from the above discussion that, in the absence of timing uncertainty, the communication signal $\sbf_c$ can be written as
\ben
\sbf_c=\Jbf\sbf,
\een
where
\[
\Jbf=
\left (
\begin{array}{ccc|cccc|ccc}
0  & \cdots & 0 & 1 & 0 & \cdots & 0 & 0 & \cdots & 0\\
0  & \ddots &\vdots & 0 & \ddots & \ddots & \vdots & 0 & \ddots & \vdots\\
\vdots  & \ddots & 0 & \vdots & \ddots & \ddots & 0 & \vdots & \ddots & 0\\
\undermat{N\times \frac{M-N}{2}}{0 & \cdots & 0} & \undermat{N\times N}{0 & \cdots & 0 & 1} & \undermat{N\times \frac{M-N}{2}}{0 & \cdots & 0} \\
\end{array}
\right ).
\]
\\
\\
Substituting \eqref{equ:ss} into the above equation, we have
\ben\label{equ:withoutsc}
\sbf_c=\Jbf\Hbf\bbf.
\een
\subsection{Timing Uncertainty}
\label{subsubsec:WithUncertainty}
As noted in \emph{Remark 2}, timing uncertainty may occur in practice due to uncertainties of the target location and IO location (IO may be placed on a moving platform, e.g., satellite). In addition, timing uncertainty may be caused by synchronization errors. All these uncertainties translate into (or can be modeled by) an uncertainty on the target location. Let $x+\Delta x$ and $y+\Delta y$ denote the true target coordinates relative to the nominal target coordinates $(x,y)$. Then, the true propagation delay difference can be expressed as
\ben
\begin{split}
\Delta\tau'=&\frac{1}{c}\big(\sqrt{(x+\Delta x-x_r)^2+(y+\Delta y-y_r)^2}\\&-\sqrt{(x+\Delta x-x_c)^2+(y+\Delta y-y_c)^2}\big).
\end{split}
\een
Since the timing uncertainty is generally unknown, the radar still uses the nominal propagation delay difference $\Delta \tau$ to perform sampling and alignment. This leads to an extra and unknown delay given by
\ben\label{equ:k}
k= \frac{\Delta\tau'-\Delta\tau}{T_s},
\een
which is assumed to be an integer. Taking this timing uncertainty into account, we can express the IO signal as
\ben\label{equ:async}
\sbf_c=\Jbf_k\Hbf\bbf,
\een
where $\Jbf_k$ is an $N\times M$ shifting matrix with the $(i,j)$-th element given by
\ben
\Jbf_k(i,j)=
\begin{cases}
  1,~i=j-\bar{k}~\text{and}~j=\bar{k}+1,\cdots,\bar{k}+N,\\
  0,~\text{otherwise},
\end{cases}
\een
where $\bar{k}=\frac{M-N}{2}-k$. It is easy to check $\Jbf_0=\Jbf$.
\subsection{Design Criteria}
\label{subsec:designcriterion}
Similar to the process in Section \ref{sec:signalmodel}, two filters $\wbf_r$ and $\wbf_c$ are employed to process the received signal and the output SINR can be expressed as
\ben\label{equ:SINR_OR}
\begin{split}
\text{SINR}_k(\sbf_r,\wbf_r,\wbf_c)&=\frac{\gamma_r\wbf_r^H\sbf_r\sbf_r^H\wbf_r}{\gamma_c\wbf_r^H\Jbf_k\Hbf\Rbf_b\Hbf^H\Jbf_k^H\wbf_r+\wbf_r^H\wbf_r}\\
&+\frac{\gamma_c\wbf_c^H\Jbf_k\Hbf\Rbf_b\Hbf^H\Jbf_k^H\wbf_c}{\gamma_r\wbf_c^H\sbf_r\sbf_r^H\wbf_c+\wbf_c^H\wbf_c},
\end{split}
\een
where $\Rbf_b=\mathbb{E}\{\bbf\bbf^H\}$. Without loss of generality, we assume $\Rbf_b=\Ibf$.

When there is no timing uncertainty, the design problem can be formulated as maximizing the $\text{SINR}_k$ in \eqref{equ:SINR_OR} with $k=0$. On the other hand, with timing uncertainty, $k$ is unknown. In this case, we consider a range of possible delays, $k=-K,\cdots,K$, where $K$ denotes an upper bound on the timing uncertainty. Specifically, we propose two design approaches to cope with timing uncertainty. The first is a max-min (MM) based approach which considers the worst-case SINR among the $2K+1$ positions as the figure of merit (to be optimized). The second is a weighted-sum (WS) based approach which employs a set of weights to form a weighted sum of the SINR at each location as a merit. The latter approach is useful in applications where we have some prior knowledge about the timing uncertainty.
\subsubsection{Max-Min} For the MM formulation, we maximize the worst-case SINR over the unknown timing uncertainty under a power constraint for the radar:
\begin{subequations}\label{equ:mm}
\begin{gather}
\label{equ:mm_cost}
\max\limits_{\sbf_r,~\wbf_r,~\wbf_c}~~\min\limits_{k\in\mathbb{N}_K}~~\text{SINR}_k(\sbf_r,\wbf_r,\wbf_c)
\\
\label{equ:mm_const}
\text{s.t.}~~\sbf_r^H\sbf_r\leq P_r.
\end{gather}
\end{subequations}
where the cost function $\text{SINR}_k(\sbf_r,\wbf_r,\wbf_c)$ is defined in \eqref{equ:SINR_OR} and \eqref{equ:mm_const} represents a power constraint for the radar.
\subsubsection{Weighted-Sum} The WS formulation aims to maximize the weighted sum of the SINR over different delays:
\begin{subequations}\label{equ:ws}
\begin{gather}
\max\limits_{\sbf_r,~\wbf_r,~\wbf_c}~~\frac{1}{2K+1}\sum_{k=-K}^{K}u_k\text{SINR}_k(\sbf_r,\wbf_r,\wbf_c)
\\
\text{s.t.}~~\sbf_r^H\sbf_r\leq P_r,
\end{gather}
\end{subequations}
where $\{u_k\}_{k=-K}^K$ are real-valued weights for the corresponding SINR. To simplify the notation, the scaling factor $\frac{1}{2K+1}$ will be ignored in the following derivation.

\emph{Remark 4}: It is easy to see that the MM and WS design criteria, which are proposed to deal with timing uncertainty, become identical to each other when $K=0$, i.e., when there is no timing uncertainty. Therefore, the design with no timing uncertainty is a special case of the one with timing uncertainty.

\emph{Remark 5}: The above formulation assumes
    knowledge of the channel SNR $\gamma_r$ \eqref{equ:gammar_r} and
    $\gamma_c$ \eqref{equ:gammar_c}. Although the target RCS $\rho$ is
    generally unknown prior to target detection, the channel
    statistics $\mathbb{E}\{\vert h_r\vert^2\}$ and $\mathbb{E}\{\vert
    h_c\vert^2\}$ are related to the locations of the Tx, Rx, and
    range bin, as well as the transmit power, which can be learned
    through training and calibration. We may assume the RCS is also
    known, since the design can be thought of as based on a nominal
    target, i.e., the smallest target that is detectable for a given
    range, which is often employed in radar design
    \cite{Blake90}. Meanwhile, since interference dominates noise in
    the considered co-channel scenario, the noise related terms,
    $\wbf_r^H\wbf_r$ and $\wbf_c^H\wbf_c$, in the denominators of
    \eqref{equ:SINR_OR} are negligible compared to their interference
    counterparts (the first term of each denominator). In such cases,
    the knowledge of the RCS is no longer necessary since $\rho$
    cancels out in the SINR expression \eqref{equ:SINR_OR}.

\section{Proposed Solutions}
\label{sec:proposedmethod}
In this section, we first develop solutions for the MM and WS optimization problems, i.e., \eqref{equ:mm} and \eqref{equ:ws}, in the presence of timing uncertainty, and then simplify the solutions to the case when no timing uncertainty is present. Both \eqref{equ:mm} and \eqref{equ:ws} are nonconvex problems without closed-form solutions. Thus, iterative processes are employed to solve these problems.

\subsection{Max-Min Design}
\label{subsec:MMformulation}
In this subsection, we introduce a sequential optimization procedure to solve the max-min problem \eqref{equ:mm} by iteratively optimizing the worst-case SINR with respect to (w.r.t.) the active radar waveform $\sbf_r$ and the receive filters $\wbf_r$ and $\wbf_c$. Specifically, by fixing $\sbf_r$ and $\wbf_c$ to the values obtained from the $\ell$-th iteration, $\sbf_r^{(\ell)}$ and $\wbf_c^{(\ell)}$, we can write \eqref{equ:mm} as
\ben\label{equ:mmwr}
\max\limits_{\wbf_r}~\min\limits_{k\in\mathbb{N}_K}~\big(\text{SINR}_{r,k}(\sbf_r^{(\ell)},\wbf_r)+\text{SINR}_{c,k}(\sbf_r^{(\ell)},\wbf_c^{(\ell)})\big),
\een
where only $\text{SINR}_{r,k}$ depends on $\wbf_r$ and is given by
\ben
\text{SINR}_{r,k}(\sbf_r^{(\ell)},\wbf_r)=\frac{\gamma_r\wbf_r^H\sbf_r^{(\ell)}(\sbf_r^{(\ell)})^H\wbf_r}{\gamma_c\wbf_r^H\Jbf_k\Hbf\Hbf^H\Jbf_k^H\wbf_r+\wbf_r^H\wbf_r},
\een
and the second SINR term is a constant w.r.t. $\wbf_r$ but depends on $k$:
\ben
\text{SINR}_{c,k}\triangleq\frac{\gamma_c(\wbf_c^{(\ell)})^H\Jbf_k\Hbf\Hbf^H\Jbf_k^H\wbf_c^{(\ell)}}{\gamma_r(\wbf_c^{(\ell)})^H\sbf_r^{(\ell)}(\sbf_r^{(\ell)})^H\wbf_c^{(\ell)}+(\wbf_c^{(\ell)})^H\wbf_c^{(\ell)}}.
\een
The optimization problem in \eqref{equ:mmwr} is still nonconvex. In the following, we use the semidefinite relaxation (SDR) \cite{LuoSo10} to convert \eqref{equ:mmwr} to a fractional programming problem by dropping the rank-one constraint. Let $\Wbf_r=\wbf_r\wbf_r^H$ and $\Sbf_r^{(\ell)}=\sbf_r^{(\ell)}(\sbf_r^{(\ell)})^H$. The SDR of \eqref{equ:mmwr} can be expressed as
\ben\label{equ:mmwrSDR}
\max\limits_{\Wbf_r\succeq\mathbf{0}}~~\min\limits_{k\in\mathbb{N}_K}~~\Big(\frac{\gamma_r\tr\big(\Wbf_r\Sbf_r^{(\ell)}\big)}{\tr\big(\Wbf_r(\gamma_c\Jbf_k\Hbf\Hbf^H\Jbf_k^H+\Ibf)\big)}+\text{SINR}_{c,k}\Big).
\een
Clearly, \eqref{equ:mmwrSDR} can be represented in the following generalized form as
\ben\label{equ:standardFP}
\max\limits_{\Wbf\in\Xcal}~\min\limits_{k\in\mathbb{N}_K}~\Big(\frac{\bar{f}_k(\Wbf)}{g_k(\Wbf)}+\kappa_k\Big),
\een
where $\Xcal$ is a convex set of positive semi-definite matrices, $\kappa_k$ is a constant independent of the variable $\Wbf$ but dependent on $k$, $\bar{f}_k(\Wbf)$ is a non-negative concave/affine function, and $g_k(\Wbf)$ is a positive convex/affine function. For our problem, $\bar{f}_i(\Wbf)$ represents the numerator of the ratio in \eqref{equ:mmwrSDR}, $g_k(\Wbf)$ corresponds to the denominator, and $\kappa_k$ is $\text{SINR}_{c,k}$. Problem \eqref{equ:standardFP} can be rewritten as a fractional programming problem and solved by the Dinkelbach algorithm in polynomial time \cite{Dinkelbach67}. For completeness, we summarize the process of solving problem \eqref{equ:standardFP} in $\textbf{Algorithm~\ref{alg:GDA}}$. The $\Wbf^{(\ell_1+1)}$ in Step 2) of $\textbf{Algorithm~\ref{alg:GDA}}$ can be obtained by solving the following convex optimization problem:
\begin{subequations}\label{equ:cvxproblem}
\ben
\Wbf^{(\ell_1+1)}=\arg \max\limits_{\Wbf\in\Xcal}~~\eta
\een
\ben
\text{s.t.}~~f_k(\Wbf)-\lambda^{(\ell_1)}g_k(\Wbf)\geq \eta,\forall k\in\mathbb{N}_K,
\een
\end{subequations}
where $\eta$ is an auxiliary variable. The convex problem \eqref{equ:cvxproblem} can be solved by standard numerical solvers, e.g., CVX \cite{cvxBoyd2014}. Thus, an optimal solution $\Wbf_r^{(\ell+1)}$ of \eqref{equ:mmwrSDR} can be obtained through $\textbf{Algorithm~\ref{alg:GDA}}$.
\begin{algorithm}
\caption{Generalized Dinkelbach's Algorithm}
\begin{algorithmic}
\label{alg:GDA}
\STATE \textbf{Input:} Known parameters that define problem \eqref{equ:standardFP}
\STATE  \textbf{Initialization:} Initialize $\lambda^{(0)}=0$. Set the error tolerance $\epsilon$ and iteration index $\ell_1=0$. Let $f_k(\Wbf)=\bar{f}_k(\Wbf)+\kappa_kg_k(\Wbf)$.
\\
\REPEAT
\STATE
\begin{enumerate}
\item $\Wbf^{(\ell_1+1)}=\arg~\max\limits_{\Wbf\in\Xcal}\min\limits_{k\in\mathbb{N}_K}~\{f_k(\Wbf)-\lambda^{(\ell_1)}g_k(\Wbf)\}$
\item Update $\lambda^{(\ell_1+1)}=\min\limits_{k\in\mathbb{N}_K}~\frac{f_k(\Wbf^{(\ell_1+1)})}{g_k(\Wbf^{(\ell_1+1)})}$
\item Let $\epsilon_0=\min\limits_{k\in\mathbb{N}_K}~\{f_k(\Wbf^{(\ell_1+1)})-\lambda^{(\ell_1)}g_k(\Wbf^{(\ell_1+1)})\}.$
\item Set $\ell_1=\ell_1+1$
\end{enumerate}
\UNTIL $\epsilon_0\leq\epsilon$.
\RETURN $\Wbf=\Wbf^{(\ell_1+1)}$.
\end{algorithmic}
\end{algorithm}

When employing the SDR approach, one has to convert the optimal solution $\Wbf_r^{(\ell+1)}$ to \eqref{equ:mmwrSDR} into a feasible solution $\wbf_r^{(\ell+1)}$ to \eqref{equ:mmwr}. We can use the randomization method to obtain a solution $\wbf_r^{(\ell+1)}$ from $\Wbf_r^{(\ell+1)}$ \cite{LuoSo10}. Note that, during the $(\ell_1+1)$-st iteration of $\textbf{Algorithm~\ref{alg:GDA}}$, the following convex optimization problem is solved
\ben\label{equ:wrD}
\max\limits_{\Wbf_r\succeq\mathbf{0}}~~\min\limits_{k\in\mathbb{N}_K}~\big(f_k(\Wbf_r)-\lambda^{(\ell_1)}g_k(\Wbf_r)\big),
\een
where the cost function $f_k(\Wbf_r)-\lambda^{(\ell_1)}g_k(\Wbf_r)$ has following form
\begin{align}
&\gamma_r\tr\big(\Wbf_r\Sbf_r^{(\ell)}\big)+\text{SINR}_{c,k}\tr\big(\Wbf_r(\gamma_c\Jbf_k\Hbf\Hbf^H\Jbf_k^H+\Ibf)\big)\notag\\
 &-\lambda^{(\ell_1)}\tr\big(\Wbf_r(\gamma_c\Jbf_k\Hbf\Hbf^H\Jbf_k^H+\Ibf)\big).
\end{align}Consider a random vector $\xibf$ with zeros mean and covariance matrix $\mathbb{E}\{\xibf\xibf^H\}$, i.e., $\xibf\sim\Ncal(\mathbf{0},\mathbb{E}\{\xibf\xibf^H\})$. It is clear that \eqref{equ:wrD} is equivalent to the following stochastic optimization problem:
\ben\label{equ:mmwrstochastic}
\max\limits_{\mathbb{E}\{\xibf\xibf^H\})\succeq\mathbf{0}}~~\min\limits_{k\in\mathbb{N}_K}~~F_k(\xibf),
\een
where
\ben
\begin{split}
&F_k(\xibf)=\gamma_r\mathbb{E}\big\{\xibf^H\Sbf_r^{(\ell)}\xibf\big\}\\&+\text{SINR}_{c,k}(\sbf_r^{(\ell)},\wbf_c^{(\ell)})\mathbb{E}\big\{\xibf^H(\gamma_c\Jbf_k\Hbf\Hbf^H\Jbf_k^H+\Ibf)\xibf\big\}\\
&-\lambda^{(\ell_1)}\mathbb{E}\big\{\xibf(\gamma_c\Jbf_k\Hbf\Hbf^H\Jbf_k^H+\Ibf)\xibf\big\}.
\end{split}
\een
Hence, the stochastic interpretation of the SDR in \eqref{equ:mmwrstochastic} allows us to obtain an approximate rank-one solution to \eqref{equ:mmwrSDR}. Specifically, given $\Wbf_r^{(\ell+1)}$, we can generate a set of independent and identically distributed Gaussian random vectors $\xibf_i\sim\Ncal(\mathbf{0},\Wbf_r^{(\ell+1)})$, $i=1,\cdots,Q$, where $Q$ is the number of randomization trials. Then, a rank-one solution is obtained as
\ben\label{equ:randomization}
\wbf_r^{(\ell+1)}=\arg~\max\limits_{\xibf_i}~\min\limits_{k\in\mathbb{N}_K}~\big(\text{SINR}_{r,k}(\sbf_r^{(\ell)},\xibf_i)+\text{SINR}_{c,k}\big).
\een

Next, we find $\Wbf_c=\wbf_c\wbf_c^H$ by fixing $\Sbf_r$ and $\Wbf_r$ to values obtained from the latest updates, $\Sbf_r^{(\ell)}=\sbf_r^{(\ell)}(\sbf_r^{(\ell)})^H$ and $\Wbf_r^{(\ell+1)}=\wbf_r^{(\ell+1)}(\wbf_r^{(\ell+1)})^H$, in which case the optimization problem \eqref{equ:mm} becomes
\ben\label{equ:mmwcSDR}
\max\limits_{\Wbf_c\succeq\mathbf{0}}\min\limits_{k\in\mathbb{N}_K}\Big(\frac{\gamma_c\tr\big(\Wbf_c\Jbf_k\Hbf\Hbf^H\Jbf_k^H\big)}{\tr\big(\Wbf_c(\gamma_r\Sbf_r^{(\ell)}+\Ibf)\big)}+G_{r,k}(\Wbf_r^{(\ell+1)},\Sbf_r^{(\ell)})\Big),
\een
where the second SINR is independent of $\Wbf_c$:
\ben
G_{r,k}(\Wbf_r^{(\ell+1)},\Sbf_r^{(\ell)})=\frac{\gamma_r\tr\big(\Wbf_r^{(\ell+1)}\Sbf_r^{(\ell)}\big)}{\tr\big(\Wbf_r^{(\ell+1)}(\gamma_c\Jbf_k\Hbf\Hbf^H\Jbf_k^H+\Ibf)\big)}.
\een
It is easy to see that $\textbf{Algorithm~\ref{alg:GDA}}$ along with the randomization method can be used to solve the above problem and find $\Wbf_c^{(\ell+1)}$ and $\wbf_c^{(\ell+1)}$.

Finally, we find $\Sbf_r=\sbf_r\sbf_r^H$ by fixing $\Wbf_r$ and $\Wbf_c$ to $\Wbf_r^{(\ell+1)}$ and $\Wbf_c^{(\ell+1)}=\wbf_c^{(\ell+1)}(\wbf_c^{(\ell+1)})^H$. With the help of SDR, the optimization problem \eqref{equ:mm} can be simplified as:
\begin{subequations}\label{equ:scpsr}
\begin{gather}
\max\limits_{\Sbf_r\succeq\mathbf{0}}~\min\limits_{k\in\mathbb{N}_K}~\big(F_{r,k}(\Sbf_r,\Wbf_r^{(\ell+1)})+F_{c,k}(\Sbf_r,\Wbf_c^{(\ell+1)})\big)
\\
\label{equ:powerconstraint}
\text{s.t.}~~\tr(\Sbf_r)\leq P_r,
\end{gather}
\end{subequations}
where
\ben
\label{equ:F_rk}
F_{r,k}(\Sbf_r,\Wbf_r^{(\ell+1)})=\frac{\gamma_r\tr\big((\Wbf_r^{\ell+1)}\Sbf_r\big)}{\tr\big(\Wbf_r^{(\ell+1)}(\gamma_c\Jbf_k\Hbf\Hbf^H\Jbf_k^H+\Ibf)\big)},
\een
and
\ben
\label{equ:F_ck}
F_{c,k}(\Sbf_r,\Wbf_c^{(\ell+1)})=\frac{\gamma_c\tr\big(\Wbf_c^{(\ell+1)}\Jbf_k\Hbf\Hbf^H\Jbf_k^H\big)}{\tr\big(\Wbf_c^{(\ell+1)}(\gamma_r\Sbf_r+\Ibf)\big)}.
\een
Although the cost function in \eqref{equ:scpsr} is convex, maximizing a convex function is a nonconvex problem \cite{Boyd2004}. We can employ a convex relaxation based sequential convex programming (SCP) approach \cite{DinhDiehl10} to solve the problem. Specifically, let $\Sbf_r^{(\ell_2)}$ denote the solution from the $\ell_2$-th iteration of the SCP iterative process. Then, $F_{c,k}(\Sbf_r,\Wbf_c^{(\ell+1)})$ can be approximated by its first-order Taylor expansion at $\Sbf_r^{(\ell_2)}$ as \cite{PetersenPedersen2012}
\begin{align}\label{equ:gkappr}
&F_{c,k}(\Sbf_r,\Wbf_c^{(\ell+1)})\approx\widetilde{F}_{c,k}(\Sbf_r,\Wbf_c^{(\ell+1)},\Sbf_r^{(\ell_2)})\notag\\
&\triangleq\frac{\gamma_c\tr\big(\Wbf_c^{(\ell+1)}\Jbf_k\Hbf\Hbf^H\Jbf_k^H\big)}{\tr\big(\Wbf_c^{(\ell+1)}(\gamma_r\Sbf_r^{(\ell_2)}+\Ibf)\big)}-\gamma_c\tr\big(\Wbf_c^{(\ell+1)}\Jbf_k\Hbf\Hbf^H\Jbf_k^H\big)\notag\\
&\times\tr\Big(\frac{\gamma_r\Wbf_c^{(\ell+1)}(\Sbf_r-\Sbf_r^{(\ell_2)})}{\Big(\tr\big(\Wbf_c^{(\ell+1)}(\gamma_r\Sbf_r^{(\ell_2)}+\Ibf)\big)\Big)^2}\Big).
\end{align}
Therefore, we can solve the following convex optimization problem:
\begin{subequations}\label{equ:scpsrnew}
\begin{gather}
\max\limits_{\Sbf_r\succeq\mathbf{0}}\min\limits_{k\in\mathbb{N}_K}\big(F_{r,k}(\Sbf_r,\Wbf_r^{(\ell+1)})+\widetilde{F}_{c,k}(\Sbf_r,\Wbf_c^{(\ell+1)},\Sbf_r^{(\ell_2)})\big)
\\
\text{s.t.}~~\eqref{equ:powerconstraint}.
\end{gather}
\end{subequations}
The iteration of the SCP method ends when the difference of the cost function over two adjacent iterations is smaller than a tolerance $\epsilon$. The radar waveform matrix after the convergence is denoted by $\Sbf_r^{(\ell+1)}$, which can be converted into $\sbf_r^{(\ell+1)}$ by using the randomization process.

The outer iterative process w.r.t. $\ell$ is repeated until the algorithm converges, e.g., the SINR improvement is smaller than a tolerance $\epsilon$. Our proposed sequential optimization algorithm for the max-min formulation is summarized in $\textbf{Algorithm~\ref{alg:MM}}$.

Note that the computational complexity of the proposed sequential optimization algorithm ($\textbf{Algorithm~\ref{alg:MM}}$) mainly depends on the number of outer iterations $\bar{\ell}$, the numbers of inner iterations, $\bar{\ell}_1$, $\bar{\ell}_2$, and $\bar{\ell}_3$, and the number of randomization trials for the semidefinite relaxation $Q$. On one hand, a convex problem was solved inside each inner iteration with a complexity of $\mathcal{O}(N^{3.5})$ if an interior-point method is used \cite{Boyd2004}, where $\mathcal{O}(\cdot)$ denotes the Landau notation. On the other hand, the computational complexity of $Q$ randomization trials is in the order of $\mathcal{O}(QN^2)$ \cite{AubryMaio2012}. Thus, the overall complexity of the proposed alternating algorithm is $\mathcal{O}\big(\bar{\ell}(\bar{\ell}_1+\bar{\ell}_2+\bar{\ell}_3)N^{3.5}\big)+\mathcal{O}\big(Q\bar{\ell}(\bar{\ell}_1+\bar{\ell}_2+\bar{\ell}_3)N^2\big)$. Numerical simulations show that $\textbf{Algorithm~\ref{alg:MM}}$ converges with a relatively small number of inner and outer iterations, e.g., about 10 iterations. Furthermore, for the number of randomizations, we find $Q\leq100$ is generally sufficient to yield a good solution.
\begin{algorithm}
\caption{Sequential Optimization Algorithm for the \mbox{Max-Min} Formulation}
\begin{algorithmic}
\label{alg:MM}
\STATE \textbf{Input:} Channel SNRs $\gamma_r$ and $\gamma_c$, IO waveform that is used to construct matrix $\Hbf$ in \eqref{equ:Hmatrix}, total radar power $P_r$, and tolerance $\epsilon$.
\STATE \textbf{Output:} Radar waveform $\sbf_r$ and receive filters $\wbf_r$ and $\wbf_c$.\\
\STATE  \textbf{Initialization:} Initialize $\sbf_r^{(0)}$ and $\wbf_c^{(0)}$, and set iteration index $\ell=0$.\\
\REPEAT
\STATE
\begin{enumerate}

  \item Fix $\Sbf_r^{(\ell)}$ and $\Wbf_c^{(\ell)}$. Use $\textbf{Algorithm~\ref{alg:GDA}}$ and randomization \eqref{equ:randomization} to obtain $\wbf_r^{(\ell+1)}$, and update $\Wbf_r^{(\ell+1)}=\wbf_r^{(\ell+1)}(\wbf_r^{(\ell+1)})^H$.
  \item Fix $\Sbf_r^{(\ell)}$ and $\Wbf_r^{(\ell+1)}$. Use $\textbf{Algorithm~\ref{alg:GDA}}$ along with randomization to find $\wbf_c^{(\ell+1)}$, and update $\Wbf_c^{(\ell+1)}=\wbf_c^{(\ell+1)}(\wbf_c^{(\ell+1)})^H$.
  \item Initialization:$\ell_2=0$ and $\Sbf_r^{(\ell_2)}=\Sbf_r^{(\ell)}$.
  \STATE \textbf{repeat}
  \begin{enumerate}
  \item  Fix $\Wbf_r^{(\ell+1)}$ and $\Wbf_c^{(\ell+1)}$.
      \item Update the radar waveform matrix $\Sbf_r^{(\ell_2+1)}$ by solving \eqref{equ:scpsrnew}.
            \item Set $\ell_2=\ell_2+1$.
  \end{enumerate}

  \STATE \textbf{until} convergence
  \item Apply randomization to obtain $\sbf_r^{(\ell+1)}$ from $\Sbf_r^{(\ell_2+1)}$. Update $\Sbf_r^{(\ell+1)}=\sbf_r^{(\ell+1)}(\sbf_r^{(\ell+1)})^H$.
         \item Set $\ell=\ell+1$.
\end{enumerate}
\UNTIL convergence.
\RETURN $\sbf_r=\sbf_r^{(\ell+1)}$, $\wbf_r=\wbf_r^{(\ell+1)}$, and $\wbf_c=\wbf_c^{(\ell+1)}$.
\end{algorithmic}
\end{algorithm}
\subsection{Weighted-Sum Design}
\label{subsec:WSformulation}
The original WS formulation in \eqref{equ:ws} is nonconvex. However, we can decompose the problem into three subproblems by sequentially and iteratively fixing two variables and solving one variable at a time. Specifically, by fixing $\sbf_r$ and $\wbf_c$ from the $\ell$-th iteration, we have
\ben\label{equ:wrws}
\max\limits_{\wbf_r}~~\sum_{k=-K}^{K}\frac{u_k\gamma_r\wbf_r^H\sbf_r^{(\ell)}(\sbf_r^{(\ell)})^H\wbf_r}{\gamma_c\wbf_r^H\Jbf_k\Hbf\Hbf^H\Jbf_k^H\wbf_r+\wbf_r^H\wbf_r},
\een
where the constant term which is independent of $\wbf_r$ is dropped. The above quadratic optimization problem can be relaxed into an SDR form as
\ben\label{equ:sdrws}
\max\limits_{\Wbf_r\succeq\mathbf{0}}~~\sum_{k=-K}^{K}\frac{u_k\gamma_r\tr\big(\Wbf_r\Sbf_r^{(\ell)}\big)}{\tr\big(\Wbf_r(\gamma_c\Jbf_k\Hbf\Hbf^H\Jbf_k^H+\Ibf)\big)},
\een
which is a sum-of-ratio problem. Dinkelbach's algorithm in $\textbf{Algorithm~\ref{alg:GDA}}$ is no longer applicable. However, a quadratic transform based approach \cite{ShenYu18} can be used to solve the sum-of-ratio problem. Similar to Dinkelbach's algorithm, this approach introduces a set of slack variables $\lambda_k$ ($k\in\mathbb{N}_K$) to deal with the non-convexity and applys alternating optimization. Specifically, problem \eqref{equ:sdrws} is equivalent to \cite[Corollary 1]{ShenYu18}
\ben
\max\limits_{\Wbf_r\succeq\mathbf{0},\lambda_k}~~F(\Wbf_r,\lambda_k,\Sbf_r^{(\ell)}),
\een
where
\ben\label{equ:F}
\begin{split}
   F(\Wbf_r,\lambda_k,\Sbf_r^{(\ell)}) & = \sum_{k=-K}^{K}\Big(2\lambda_k\sqrt{u_k\gamma_r\tr\big(\Wbf_r\Sbf_r^{(\ell)}\big)}\\
     & -\lambda_k^2\tr\big(\Wbf_r(\gamma_c\Jbf_k\Hbf\Hbf^H\Jbf_k^H+\Ibf)\big)\Big).
\end{split}
\een
Let $\lambda_k^{(\ell_1)}$ and $\Wbf_r^{(\ell_1)}$ denote the solutions obtained from the $\ell_1$-th inner iteration. Then, $\Wbf_r^{(\ell_1+1)}$ can be updated by solving the following convex problem:
\ben\label{equ:WR}
\max\limits_{\Wbf_r\succeq\mathbf{0}}~F(\Wbf_r,\lambda_k^{(\ell_1)},\Sbf_r^{(\ell)}),
\een
where $F(\Wbf_r,\lambda_k^{(\ell_1)},\Sbf_r^{(\ell)})$ has the same form of \eqref{equ:F} by fixing $\lambda_k=\lambda_k^{(\ell_1)}$.
Similarly, $\lambda_k^{(\ell_1+1)}$ can be obtained by solving the following problem:
\ben\label{equ:lambda}
\max\limits_{\lambda_k}~F(\Wbf_r^{(\ell_1+1)},\lambda_k,\Sbf_r^{(\ell)}),
\een
which has a closed-form solution:
\ben\label{equ:lambdak}
\lambda_k^{(\ell_1+1)}=\frac{\sqrt{u_k\gamma_r\tr\big(\Wbf_r^{(\ell_1+1)}\Sbf_r^{(\ell)}\big)}}{\tr\big(\Wbf_r^{(\ell_1+1)}(\gamma_c\Jbf_k\Hbf\Hbf^H\Jbf_k^H+\Ibf)\big)}.
\een
The alternating process between \eqref{equ:WR} and \eqref{equ:lambda} ends when the improvement of the cost function over two adjacent iterations is smaller than a tolerance $\epsilon$. After convergence, the filter matrix $\Wbf_r^{(\ell+1)}$ is used to find $\wbf_r^{(\ell+1)}$ by applying randomization process.

Next, we fix $\sbf_r=\sbf_r^{(\ell)}$ and $\wbf_r=\wbf_r^{(\ell+1)}$. The optimization problem \eqref{equ:ws} w.r.t. $\wbf_c$ can be simplified by dropping the constant term:
\ben\label{equ:WC}
\max\limits_{\wbf_c}~\frac{\wbf_c^H\big(\gamma_c\sum_{k=-K}^Ku_k\Jbf_k\Hbf\Hbf^H\Jbf_k^H\big)\wbf_c}{\wbf_c^H(\gamma_r\sbf_r^{(\ell)}(\sbf_r^{(\ell)})^H+\Ibf)\wbf_c},
\een
where the cost function is a generalized Rayleigh quotient and can be reduced to a Rayleigh quotient through a transformation $\Rbf\triangleq\Cbf^{-1}\big(\gamma_c\sum_{k=-K}^Ku_k\Jbf_k\Hbf\Hbf^H\Jbf_k^H\big)\Cbf^{-H}$ where $\Cbf\Cbf^H$ is the Cholesky decomposition of the Hermitian positive-definite matrix $(\gamma_r\sbf_r^{(\ell)}(\sbf_r^{(\ell)})^H+\Ibf)$. Then, \eqref{equ:WC} is solved in closed-form, that is, $\wbf_c^{(\ell+1)}$ is the eigenvector of matrix $\Rbf$ corresponding to the largest eigenvalue.

Finally, fix $\wbf_r$ and $\wbf_c$ to the values obtained in the current iteration and let $\Wbf_r^{(\ell+1)}=\wbf_r^{(\ell+1)}(\wbf_r^{(\ell+1)})^H$ and $\Wbf_c^{(\ell+1)}=\wbf_c^{(\ell+1)}(\wbf_c^{(\ell+1)})^H$, we have
\begin{subequations}\label{equ:SR}
\ben
\max\limits_{\Sbf_r\succeq\mathbf{0}}\sum_{k=-K}^Ku_k\Big(F_{r,k}(\Sbf_r,\Wbf_r^{(\ell+1)})+F_{c,k}(\Sbf_r,\Wbf_c^{(\ell+1)})\Big)
\een
\ben\label{equ:powerc}
\text{s.t.}~\tr(\Sbf_r)\leq P_r,
\een
\end{subequations}
where $F_{r,k}(\Sbf_r,\Wbf_r^{(\ell+1)})$ and $F_{c,k}(\Sbf_r,\Wbf_c^{(\ell+1)})$ are defined in \eqref{equ:F_rk} and \eqref{equ:F_ck}, respectively. Like \eqref{equ:scpsr}, the above problem is nonconvex and can be similarly solved by the SCP iterative approach. Specifically, during the $(\ell_2+1)$-st iteration of the SCP process, the following convex optimization problem is solved:
\begin{subequations}\label{equ:ws_s_r}
\ben
\max\limits_{\Sbf_r\succeq\mathbf{0}}\sum_{k=-K}^Ku_k\Big(F_{r,k}(\Sbf_r,\Wbf_r^{(\ell+1)})+\widetilde{F}_{c,k}(\Sbf_r,\Wbf_c^{(\ell+1)},\Sbf_r^{(\ell_2)})\Big)
\een
\ben
\text{s.t.}~~\eqref{equ:powerc}.
\een
\end{subequations}
where $\widetilde{F}_{c,k}(\Sbf_r,\Wbf_c^{(\ell+1)},\Sbf_r^{(\ell_2)})$ is defined in \eqref{equ:gkappr}. Similarly, the iteration of the SCP method ends when the difference of the cost function of two adjacent iterations is smaller than a tolerance $\epsilon$. Then, the obtained $\Sbf_r^{(\ell+1)}$ is utilized to find $\sbf_r^{(\ell+1)}$ by employing the randomization.

The outer alternating process is repeated until the algorithm converges, e.g., the SINR improvement is smaller than a tolerance $\epsilon$. Our proposed solution to the weighted-sum formulation is summarized in $\textbf{Algorithm~\ref{alg:WS}}$.

Like $\textbf{Algorithm~\ref{alg:MM}}$, the complexity of $\textbf{Algorithm~\ref{alg:WS}}$ mainly depends on the number of outer iterations $\hat{\ell}$, the numbers of inner iterations, $\hat{\ell}_1$ and $\hat{\ell}_2$, and the number of randomization trials $Q$. The overall computational complexity is thus $\mathcal{O}\big(\hat{\ell}(\hat{\ell}_1+\hat{\ell}_2)N^{3.5}\big)+\mathcal{O}\big(Q\hat{\ell}(\hat{\ell}_1+\hat{\ell}_2)N^2\big)$.
\begin{algorithm}
\caption{Sequential Optimization Algorithm for the \mbox{Weighted-Sum} Formulation}
\begin{algorithmic}
\label{alg:WS}
\STATE \textbf{Input:} Channel SNRs $\gamma_r$ and $\gamma_c$, IO waveform that is used to construct matrix $\Hbf$ in \eqref{equ:Hmatrix}, total radar power $P_r$, and tolerance $\epsilon$.
\STATE \textbf{Output:} Radar waveform $\sbf_r$ and receive filters $\wbf_r$ and $\wbf_c$.\\
\STATE  \textbf{Initialization:} Initialize $\sbf_r^{(0)}$, $\wbf_c^{(0)}$, and $\wbf_r^{(0)}$, and set iteration index $\ell=0$.\\
\REPEAT
\STATE
\begin{enumerate}

\item Initialization: $\ell_1=0$ and $\Wbf_r^{(\ell_1)}=\Wbf_r^{(\ell)}$.
\STATE \textbf{repeat}
\begin{enumerate}
\item Solve \eqref{equ:WR} for $\Wbf_r^{(\ell_1+1)}$ with fixed $\lambda_k^{(\ell_1)}$, $\Sbf_r^{(\ell)}$, and $\Wbf_c^{(\ell)}$.
\item Update $\lambda_k^{(\ell_1+1)}$ by using \eqref{equ:lambdak}.
\item $\ell_1=\ell_1+1$.
\end{enumerate}
\STATE \textbf{until} convergence.
\item Apply randomization to obtain $\wbf_r^{(\ell+1)}$ and update $\Wbf_r^{(\ell+1)}=\wbf_r^{(\ell+1)}(\wbf_r^{(\ell+1)})^H$.
  \item Obtain $\wbf_c^{(\ell+1)}$ by solving \eqref{equ:WC}.
    \item Initialization:$\ell_2=0$ and $\Sbf_r^{(\ell_2)}=\Sbf_r^{(\ell)}$.
  \STATE \textbf{repeat}
  \begin{enumerate}
  \item  Fix $\Wbf_r^{(\ell+1)}$ and $\Wbf_c^{(\ell+1)}=\wbf_c^{(\ell+1)}(\wbf_c^{(\ell+1)})^H$.
  \item Solve \eqref{equ:ws_s_r} for radar waveform matrix $\Sbf_r^{(\ell_2+1)}$.
  \item Set $\ell_2=\ell_2+1$.
  \end{enumerate}

  \STATE \textbf{until} convergence
  \item Find $\sbf_r^{(\ell+1)}$ by applying randomization and update $\Sbf_r^{(\ell+1)}=\sbf_r^{(\ell+1)}(\sbf_r^{(\ell+1)})^H$.
\item Set $\ell=\ell+1$.
\end{enumerate}
\UNTIL convergence.
\RETURN $\sbf_r=\sbf_r^{(\ell+1)}$, $\wbf_r=\wbf_r^{(\ell+1)}$, and $\wbf_c=\wbf_c^{(\ell+1)}$.
\end{algorithmic}
\end{algorithm}

\subsection{Simplified Solution with No Timing Uncertainty}
\label{subsec:proposedmethodwithout}
When there is no timing uncertainty, i.e., $K=0$, both the max-min design \eqref{equ:mm} and the weighted-sum design \eqref{equ:ws} become
\begin{subequations}\label{equ:nouncertainty}
\begin{gather}
\label{equ:no_cost}
\max\limits_{\sbf_r,~\wbf_r,~\wbf_c}~\text{SINR}_0(\sbf_r,\wbf_r,\wbf_c)
\\
\label{equ:no_const}
\text{s.t.}~~\sbf_r^H\sbf_r\leq P_r,
\end{gather}
\end{subequations}
where $\text{SINR}_0(\sbf_r,\wbf_r,\wbf_c)$ is defined in \eqref{equ:SINR_OR} with $k=0$. Although there is still no closed-form solution for \eqref{equ:nouncertainty}, the solution can be simplified when compared with the MM and WS designs. Specifically, by fixing $\sbf_r$ and $\wbf_c$ to the values obtained from the $\ell$-th iteration as $\sbf_r^{(\ell)}$ and $\wbf_c^{(\ell)}$, \eqref{equ:nouncertainty} reduces to the following optimization problem by dropping the constant term:
\ben\label{equ:MVDRwr}
\max\limits_{\wbf_r}~~\frac{\gamma_r\wbf_r^H\sbf_r^{(\ell)}(\sbf_r^{(\ell)})^H\wbf_r}{\wbf_r^H(\gamma_c\Jbf\Hbf\Hbf^H\Jbf^H+\Ibf)\wbf_r},
\een
which has an analytical solution
\ben\label{equ:soluwrwc}
\wbf_r^{(\ell+1)}=\frac{(\gamma_c\Jbf\Hbf\Hbf^H\Jbf^H+\Ibf)^{-1}\sbf_r^{(\ell)}}{(\sbf_r^{(\ell)})^H(\gamma_c\Jbf\Hbf\Hbf^H\Jbf^H+\Ibf)^{-1}\sbf_r^{(\ell)}}.
\een
The above closed-form solution not only bypasses the iterative procedure employed in solving \eqref{equ:mmwr} and \eqref{equ:wrws} but also saves the randomization process required to convert the SDR solution $\Wbf_r^{(\ell+1)}$ to $\wbf_r^{(\ell+1)}$.

As for the optimization w.r.t. $\wbf_c$, a similar procedure used to solve \eqref{equ:WC} can be employed by setting $K=0$, which also leads to a closed-form solution. The iteration and randomization procedures are again omitted when compared with solving \eqref{equ:mmwcSDR}. However, the SCP method and randomization process are still required to find $\sbf_r^{(\ell+1)}$ by fixing $\Wbf_r^{(\ell+1)}$ and $\Wbf_c^{(\ell+1)}$.

\section{Numerical Simulations}
\label{sec:simulationresults}
In this section, numerical examples are provided to illustrate the effectiveness of the proposed schemes and methods. We first demonstrate the performance of the proposed hybrid design for scenarios without and with timing uncertainty, respectively. Then, an application of the proposed designs to radar target detection is considered.

The simulation setup consists of a radar and a communication transmitter operating in the same frequency band. In the simulation, the number of samples in each radar pulse is $N=16$, the number of communication symbols in one radar PRI is $L=10$, and the number of samples per symbol is $P=8$. The communication signal $s_c(t)$ in \eqref{equ:ldm} is a binary phase shift keying (BPSK) signal, where the symbol waveform $g(t)$ is a raised cosine function with a rolloff factor of 0.22 truncated to a duration of $I=2$ symbol intervals. The noise variance at the radar receiver is $\sigma^2=1$. The radar and communication waveforms in \eqref{equ:sync} are normalized so that they have unit energy, i.e., $\sbf_r^H\sbf_r=\sbf_c^H\sbf_c=1$.  As such, the signal strength of the active and passive path is controlled by the channel SNR $\gamma_r$ \eqref{equ:gammar_r} and $\gamma_c$ \eqref{equ:gammar_c}, respectively.

For comparison, we consider the following 4 different system configurations:
\begin{itemize}
\item \textbf{hybrid-TxRx}: This refers to the proposed hybrid system with joint design of the radar transmit waveform and receive filters discussed in Section \ref{sec:proposedmethod}, which covers two cases with or without timing uncertainty, and two different designs (MM and WS) for the case with timing uncertainty. The specific case/design is identified in each individual experiment.
\item \textbf{hybrid-Rx}: This is a simplified version of the proposed hybrid system, which only optimizes the receiver filters $\wbf_r$ and $\wbf_c$, while fixing the radar transmit waveform $\sbf_r$. In the simulation, the radar waveform is generated as a random binary sequence of length $N$. This configuration is included in comparison to illustrate the benefit of joint Tx and Rx design versus Rx-only design in hybrid radar.
\item \textbf{active-only} or \textbf{passive-only}: These two conventional radar configurations employ either the active path or passive path for sensing. While there is no cross interference in these configurations, they cannot benefit from the energy and diversity gains offered by exploiting both the active and passive illuminating sources.
\end{itemize}

For the hybrid schemes, we have the following parameters. The weights of the WS formulation are $u_k=1$, $k=-K,\cdots,K$, the radar waveform is initialized with a randomly generated binary sequence, the number of randomization trials is $Q=200$, and the convergence tolerance is $\epsilon=0.01$.
\subsection{With No Timing Uncertainty}
\begin{figure}[tb]
\centering
\includegraphics[width=3in]{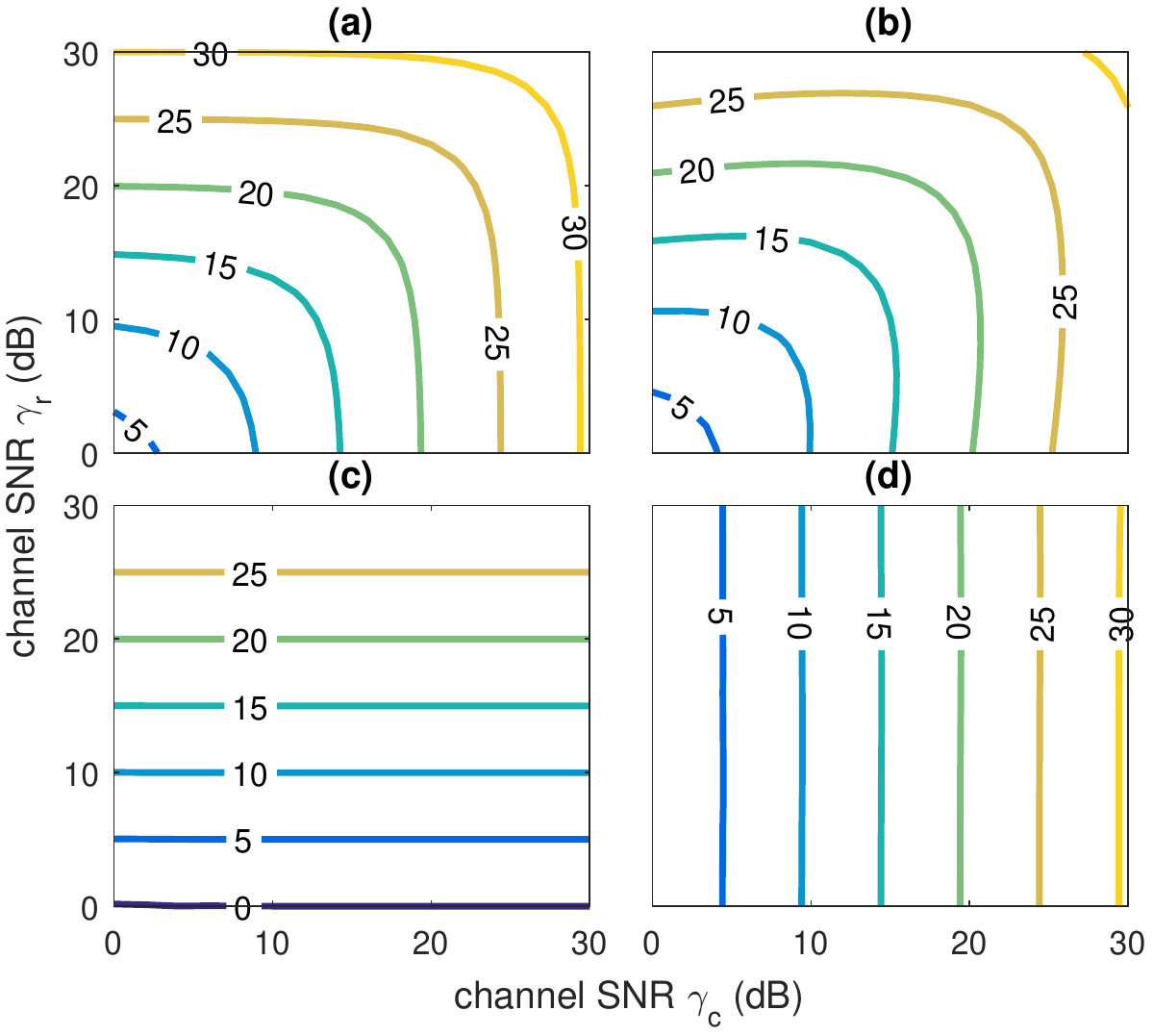}
\caption{Contour plot of the output SINR versus the channel SNR $\gamma_r$ and $\gamma_c$. (a) Hybrid-TxRx; (b) hybrid-Rx; (c) active-only; (d) passive-only.}
\label{fgr:radarpower}
\end{figure}
We first consider the performance of the hybrid schemes without timing uncertainty, i.e., $K=0$, in which case the MM and WS designs coincide. Fig.\,\ref{fgr:radarpower} shows the contour plot of the output SINR \eqref{equ:SINR_OR} versus the channel SNRs $\gamma_r$ and $\gamma_c$. Each plot contains the isolines of the output SINR with a stepsize of 5 dB. For the active-only or passive-only system, the contours are horizontal or vertical lines, as their output SINR only depends on the signal strength of a single path, determined by the transmit power, Tx-target-Rx distance, target radar cross section (RCS), etc. In contrast, the hybrid schemes can leverage the illuminations from both the active and passive paths, leading to more flexible and energy-efficient sensing solutions. For example, suppose a sensing task requires the system to provide an output $\text{SINR}=25$ dB. The hybrid schemes allow the radar Tx to significantly lower its transmit power (e.g., to close to 0 dB), if there is a strong passive source available. As the strength of the passive source decreases, the hybrid system traces the 25-dB contour line counter-clockwise and increases the transmit power of the active source. A comparison between the two hybrid schemes shows that hybrid-TxRx offers an additional gain in the output SINR over hybrid-Rx due to a better interference handling ability of the former.
\subsection{With Timing Uncertainty}

\begin{figure}[tb]
\centering
\includegraphics[width=2.5in]{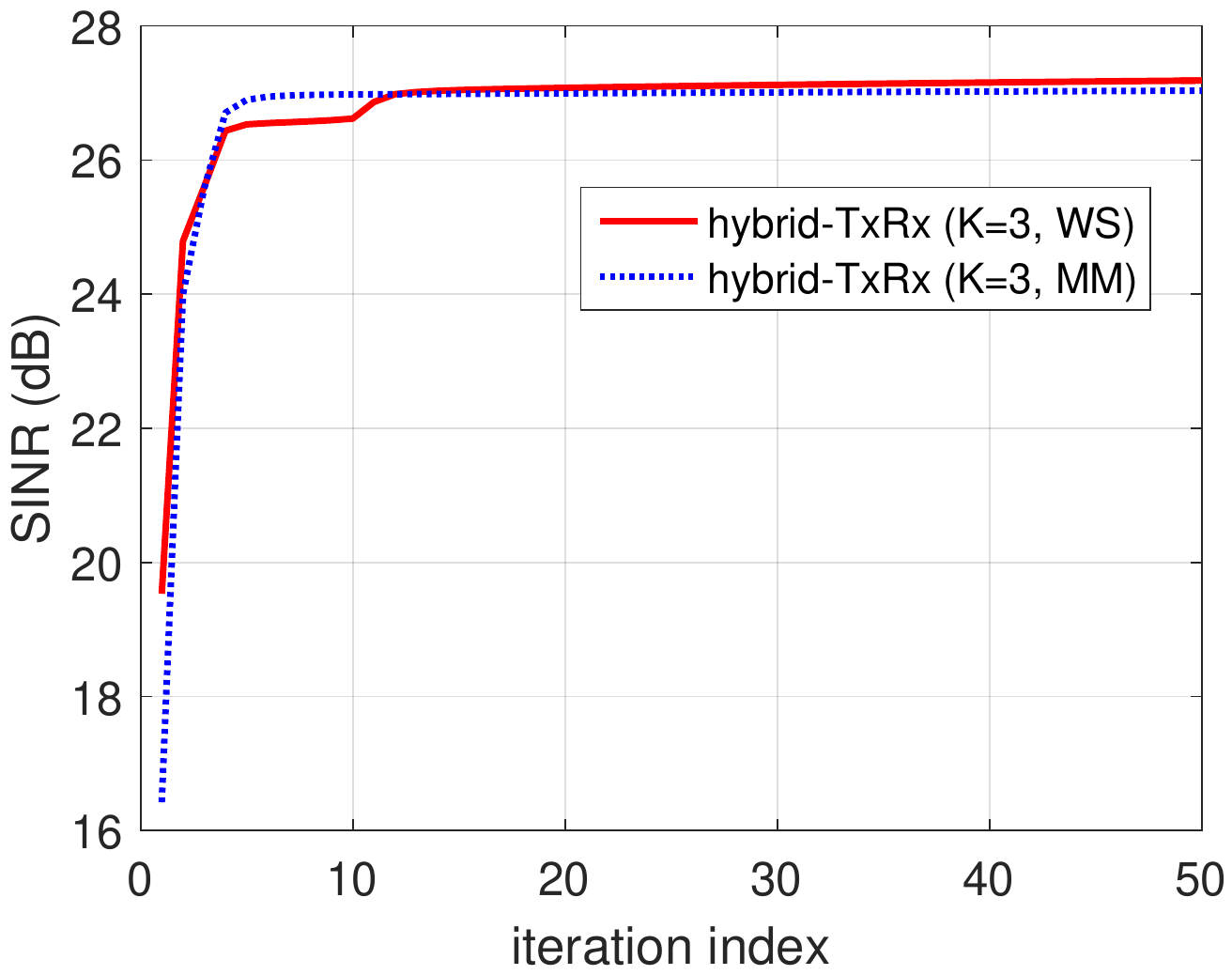}
\caption{Convergence behavior of the proposed sequential optimization algorithms when $K = 3$.}
\label{fgr:convergence}
\end{figure}
Here we only consider hybrid-TxRx and hybrid-Rx here since the timing uncertainty only affects the hybrid schemes. The channel SNRs are fixed to $\gamma_r=\gamma_c=25\text{ dB}$. We first examine the convergence of the proposed algorithms. Fig.\,\ref{fgr:convergence} depicts the output SINR of the hybrid-based design versus the number of iterations, for the MM and WS designs when $K=3$. It is seen that both designs converge quickly.

\begin{figure}[t]
\centering
\includegraphics[width=2.5in]{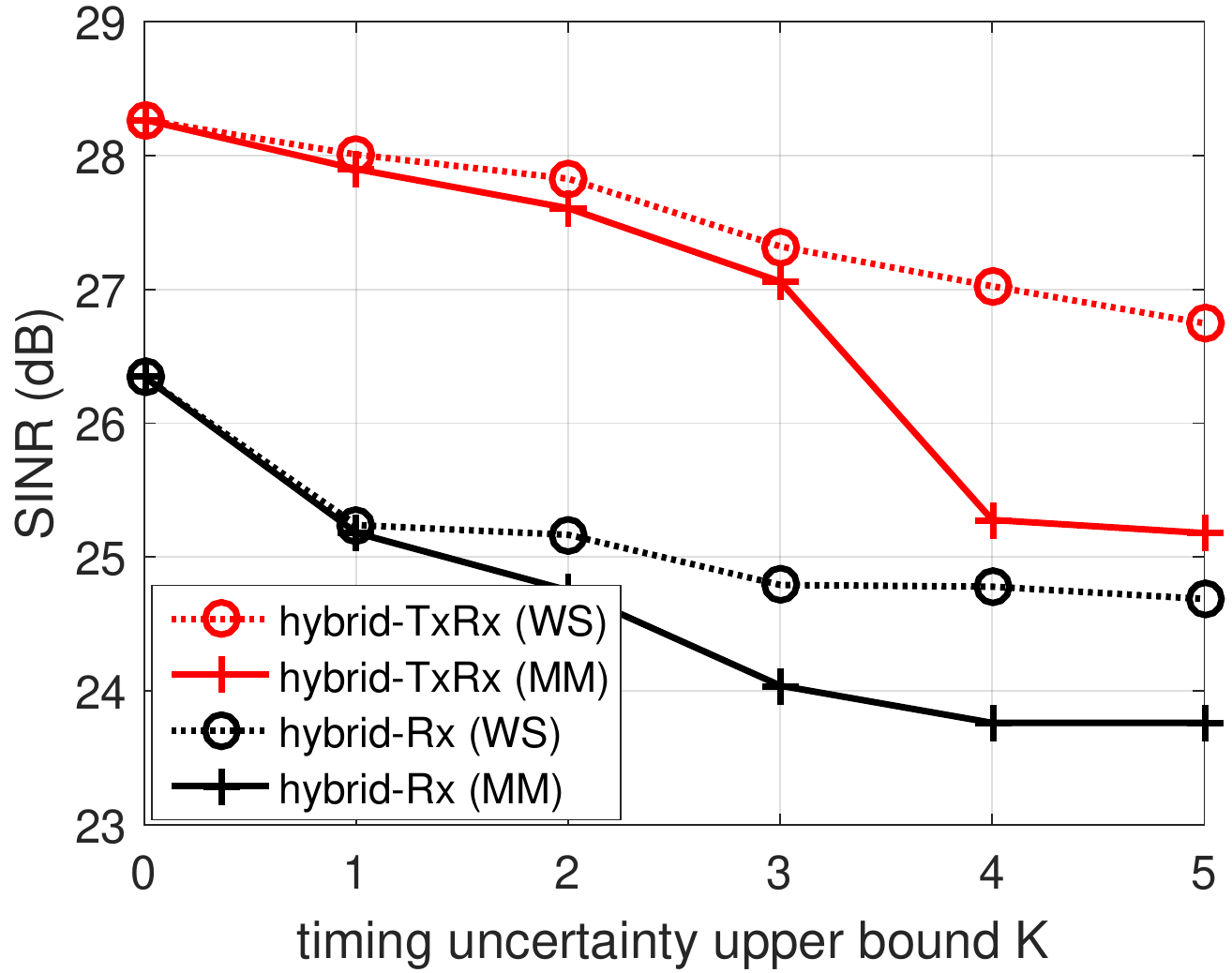}
\caption{Output SINR of the proposed hybrid designs versus the timing uncertainty bound $K$.}
\label{fgr:maxdelayuncertainty}
\end{figure}

Next, we evaluate the effects of the upper bound $K$ of the timing uncertainty. Fig.\,\ref{fgr:maxdelayuncertainty} shows the output SINR versus $K$. It is seen that as the timing uncertainty $K$ increases, the output SINR for all hybrid designs decreases. This is because a larger $K$ implies that the radar waveform and received filters under design have to deal with more mismatched cases between the radar and communication signals received by the radar Rx, which in turn imposes more difficult constraints on the design and reduces the solution space and the output SINR. Meanwhile, hybrid-TxRx is overall better than hybrid-Rx, as observed before in Fig.\,\ref{fgr:radarpower}. Within either the hybrid-TxRx or hybrid-Rx configurations, the WS design outperforms the MM design, especially for large $K$. This is because MM is based on the more conservative maxmin optimization.

\begin{figure}[t]
\centering
\includegraphics[width=2.5in]{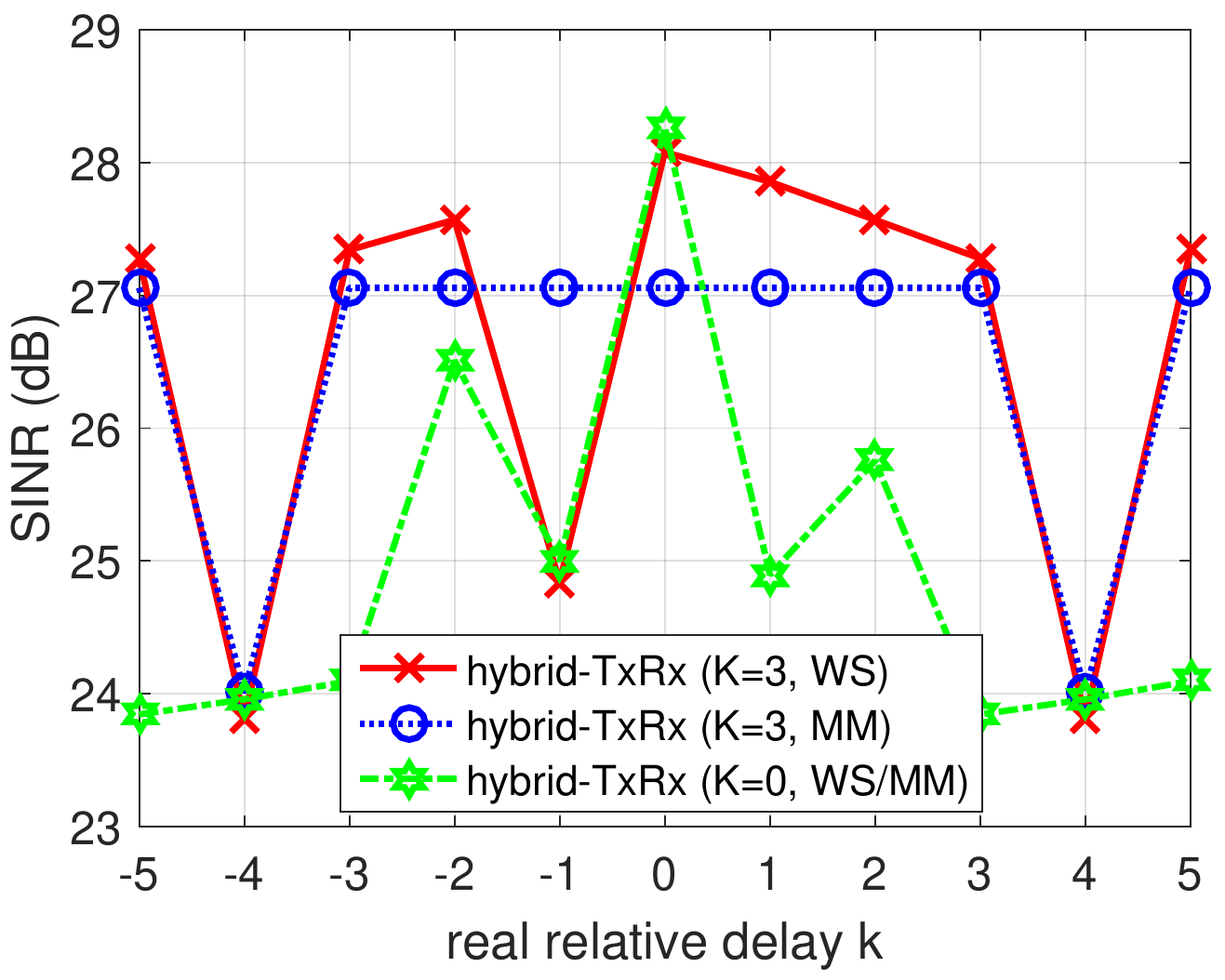}
\caption{Output SINR versus the real relative delay $k$ between the radar and communication signals.}
\label{fgr:robustness}
\end{figure}

Finally, we consider the robustness of hybrid-TxRx designed by the WS and MM approaches. Specifically, the designs are obtained by using a specific value for the timing uncertainty bound $K$, and then tested by using different values for the \emph{real} relative delay $k$ \eqref{equ:k} between the communication and radar signals. Fig.\,\ref{fgr:robustness} shows the output SINR of the WS and MM designs obtained with $K=0$ and $K=3$, respectively, versus $k$. It is observed that in the presence of timing uncertainty $k\neq0$, the designs obtained with a non-zero $K=3$ are generally better than those obtained with $K=0$, which is expected, since the timing mismatch is ignored by the latter. Interestingly, the WS design with $K=3$ incurs only a small SINR loss at $k=0$. The MM design is able to provide a uniform output SINR over the timing uncertainty interval used for the design, i.e., $-3\leq k\leq3$, although this is achieved at a slightly larger SINR loss experienced at $k=0$, when compared with the WS design.

\subsection{Detection Performance}
The previous results show the proposed hybrid designs offer higher output SINR, which is expected to yield better performance in detection than the conventional active-only or passive-only system. To see this, we consider the following hypothesis test:
\begin{subequations}
\begin{align}
\label{equ:hypothesesH0}
&H_0:~\ybf=\wbf,
\\
\label{equ:hypothesesH1}
&H_1:~\ybf=\alpha_r\sbf_r+\alpha_c\sbf_c+\wbf,
\end{align}
\end{subequations}
where under the $H_0$ hypothesis, the radar observes only noise $\wbf$, while under the $H_1$ hypothesis, the radar observation $\ybf$ contains target echoes from both the active and passive paths, as in \eqref{equ:sync}. Note that the communication waveform can be expressed as $\sbf_c=\Jbf\Hbf\bbf$ per \eqref{equ:withoutsc}. The received signal passes through two filters $\wbf_r$ and $\wbf_c$, outputting two samples $y_r$ and $y_c$, similarly as in \eqref{equ:mfout}. The two samples are combined by using the energy detector for detection \cite{HaimovichBlum08}:
\begin{equation}
|y_r|^2+|y_c|^2
\gs_{H_0}^{H_1} \zeta ,
\label{eq:lrt}
\end{equation}
where $\zeta$ is a threshold set based on a desired probability of false alarm $P_f$.
\begin{figure}[t]
\centering
\includegraphics[width=2.5in]{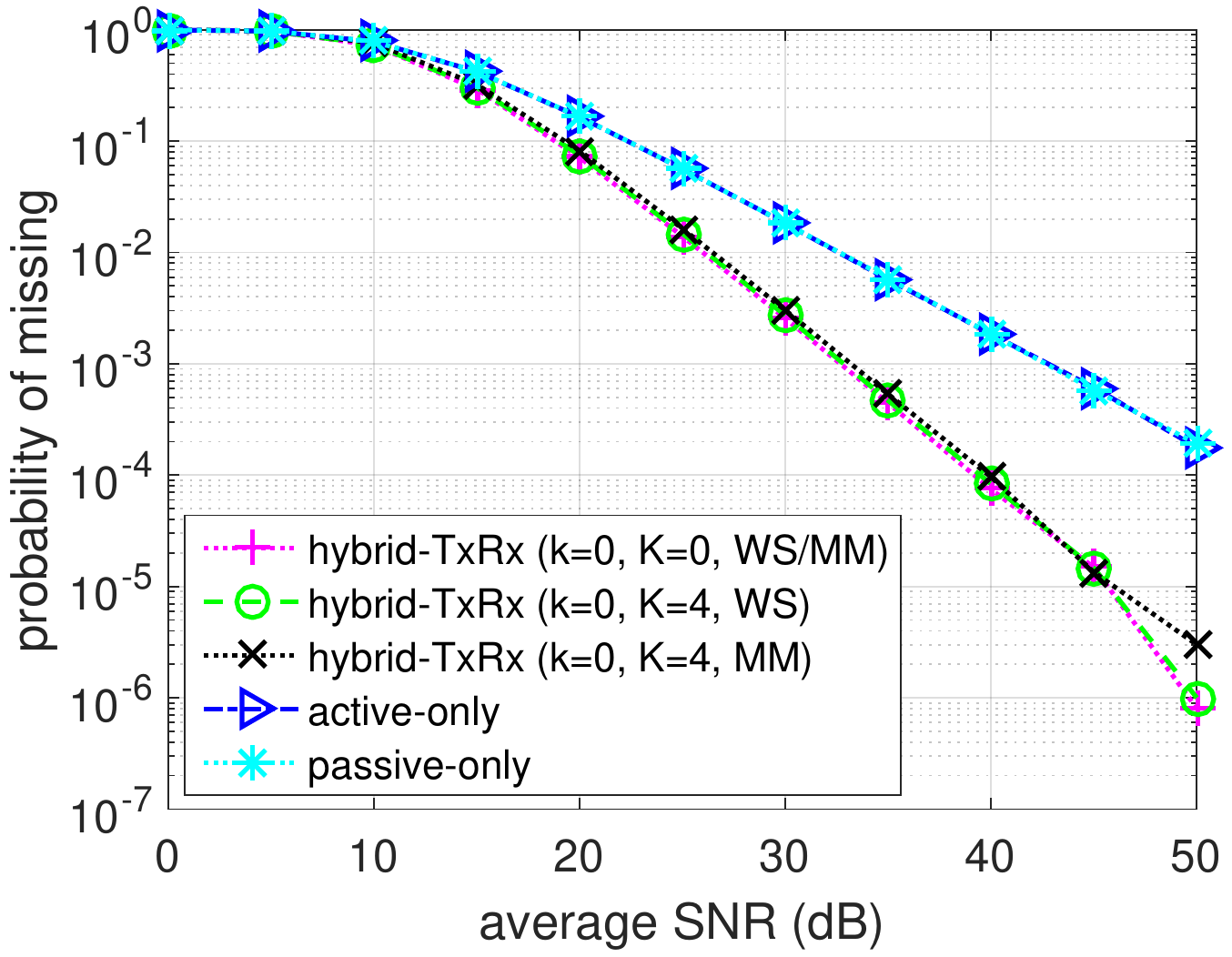}
\caption{Probability of missing versus the average SNR $\gamma$ for a given probability of false alarm $P_f$ of $10^{-4}$.}
\label{fgr:probabilityofmissing3}
\end{figure}

We consider a setup with equal channel average SNR for the active and passive paths, i.e., $\mathbb{E}\{\gamma_r\}=\mathbb{E}\{\gamma_c\}$, which corresponds to a symmetric scenario where the communication Tx and radar Tx are equally distanced from the target. The real SNR $\gamma_r$ and $\gamma_c$ are changing from one trial to another, to reflect fluctuations of the RCS of the real target. Specifically, the target RCS $\alpha_r$ and $\alpha_c$ are generated as independent complex Gaussian random variables for each trial, with zero mean and variance determined by the average SNR. In addition, the communication symbols $\bbf$ consist of random binary sequence in each trial. Our designs of the waveform and receivers are obtained by using a fixed and hypothetical value of $25\text{ dB}$ for the $\gamma_r$ and $\gamma_c$, while the other design parameters, i.e., $N$, $L$, $P$, and $I$, are identical to those used in previous examples.

We compare the detection performance of the active-only, passive-only, and the proposed hybrid-TxRx designs obtained using MM or WS, respectively. Two cases for hybrid-TxRx are considered, namely $k=0$ and $K=0$, which represents the ideal case of no timing uncertainty, and $k=0$ and $K=4$, a case with timing uncertainty.

Fig.\,\ref{fgr:probabilityofmissing3} shows the probability of missing of the various schemes versus the average SNR, when the probability of false alarm is $P_f=10^{-4}$. It is seen that the hybrid designs substantially outperform the active-only/passive-only schemes, especially at the medium-to-high SNR region. This is because the hybrid designs can leverage both the active and passive sources, and more importantly, the \emph{spatial diversity} \cite{HaimovichBlum08} created by the active and passive paths. In particular, the probability of missing for the hybrid system is lower with a steeper slope than that of the active-only and passive-only systems as the SNR increases. The steeper slope is an indicator of the diversity gain. Interestingly, it is noticed that the WS hybrid design with timing uncertainty yields nearly identical performance when compared with the ideal case of no timing uncertainty.

\section{Conclusions}
\label{sec:conclusion}
Joint design of the waveform and receivers for a hybrid active-passive radar system consisting of a monostatic active radar and a non-cooperative IO was studied by considering the location-induced timing uncertainty between the radar signal and communication signal. Two design approaches were proposed, namely a max-min based design which maximizes the worst-case SINR among all delays within a delay upper bound, and a weighted-sum based design that employs a set of weights to form a weighted sum of the SINR at each delay as a design criterion. Numerical results show that the proposed joint transmit-receive designs of the hybrid active-passive system offers significant performance gain over the conventional active-only or passive-only radar system. Future directions of interest include extensions of the current approach to cases involving multiple IO sources, joint waveform and receive designs with additional constraints, such as constant envelope, and consideration of interference to the communication system due to radar co-channel transmission.

\bibliographystyle{IEEEtran}
\bibliography{Hybrid}

\end{document}

%% file: common.tex
\newcommand{\ba}{\begin{array}}
\newcommand{\ea}{\end{array}}
\newcommand{\be}{\begin{displaymath}}
\newcommand{\ee}{\end{displaymath}}
\newcommand{\ben}{\begin{equation}}
\newcommand{\een}{\end{equation}}
\newcommand{\bena}{\begin{eqnarray}}
\newcommand{\eena}{\end{eqnarray}}
\newcommand{\beqa}{\begin{eqnarray*}}
\newcommand{\enqa}{\end{eqnarray*}}

\newcommand{\bc}{\begin{center}}
\newcommand{\ec}{\end{center}}
\newcommand{\bi}{\begin{itemize}}
\newcommand{\ei}{\end{itemize}}
\newcommand{\benu}{\begin{enumerate}}
\newcommand{\eenu}{\end{enumerate}}
\newcommand{\bdes}{\begin{description}}
\newcommand{\edes}{\end{description}}
\newcommand{\bt}{\begin{tabular}}
\newcommand{\et}{\end{tabular}}

\newcommand \xibf{\boldsymbol{\xi}}

\newcommand \Sigmabf{\boldsymbol{\Sigma}}

\newcommand \bbf{{\bf b}}

\newcommand \gbf{{\bf g}}

\newcommand \sbf{{\bf s}}

\newcommand \ubf{{\bf u}}

\newcommand \wbf{{\bf w}}

\newcommand \ybf{{\bf y}}

\newcommand \Abf{{\bf A}}

\newcommand \Cbf{{\bf C}}

\newcommand \Hbf{{\bf H}}
\newcommand \Ibf{{\bf I}}
\newcommand \Jbf{{\bf J}}

\newcommand \Rbf{{\bf R}}
\newcommand \Sbf{{\bf S}}

\newcommand \Wbf{{\bf W}}

\newcommand{\Cset}{{\mathbb C}}

\newcommand{\Ncal}{\mathcal{N}}

\newcommand{\Xcal}{\mathcal{X}}

\newcommand{\gs}{\mathop{\gtrless}\limits}

\newcommand{\circlambda}{\mbox{$\Lambda$
             \kern-.85em\raise1.5ex
             \hbox{$\scriptstyle{\circ}$}}\,}

\newcommand{\tr}{\mathop{\rm tr}}

